%% file: FADI.tex
\documentclass[modern]{aastex631}
\usepackage[titles]{tocloft}
\usepackage{newtxtext,newtxmath} 
\usepackage[USenglish]{babel}
\usepackage[T1]{fontenc}
\usepackage{comment}
\usepackage{enumitem}
\usepackage[numberedsection]{glossaries}
\usepackage{todonotes}
\usepackage{collect}
\usepackage{wrapfig}
\usepackage{bclogo}
\usepackage{environ}
\usepackage{lipsum}
\usepackage{makecell}

\setlist{noitemsep}

\graphicspath{{./}{./figures/}{./images}{./tabs}}

\newcommand{\Contact}[1]{}
\newcommand{\mail}[1]{\href{mailto:#1}{#1}}

\newcommand{\Lead}[1]{}
\newcommand{\Contributors}[2]{}

\newcommand{\midcolwid}{3in}

\providecommand{\secref}[1]{\hyperref[#1]{\autoref{#1}}}
\providecommand{\appref}[1]{\hyperref[#1]{Appendix \ref{#1}}}
\providecommand{\tabref}[1]{\hyperref[#1]{\autoref{#1}}}
\providecommand{\figref}[1]{\hyperref[#1]{\autoref{#1}}}
\providecommand{\eqnref}[1]{\hyperref[#1]{Eq.~\ref{#1}}}
\providecommand{\recref}[1]{\hyperref[#1]{REC-\ref{#1}}}

\makeatletter
\renewcommand{\section}{\newpage{} \@startsection
{section}
{1}
{0mm}
{-3.5ex \@plus -1ex \@minus -.2ex}
{0.2ex \@plus.2ex}
{\normalfont\bfseries}}
\makeatother
\input{aglossary.tex}  

\makeglossaries

\begin{document}
\pagenumbering{roman}

\title{The Future of Astronomical Data Infrastructure: \\ Meeting Report}
\shorttitle{FADI Report}

\collaboration{73}{} 
\input{authors.tex}

\maketitle

\Contact{William O'Mullane \mail{womullan@lsst.org}}

\addcontentsline{toc}{section}{Executive Summary}
\input{execsum}

\tableofcontents
\newpage
\clearpage

\pagenumbering{arabic}
\setcounter{page}{1}

\input{intro}
\input{problem}

\input{ecosys_short}

\input{scope}
\newpage
\input{casestudies}

\input{model1}
\newpage
\input{model2}

\newpage
\input{model3}

\newpage
\input{model4}
\newpage
\input{conclusions}
\clearpage

\input{acknowledgments}

\clearpage
\appendix
\input{dives}
\input{hopes}
\input{costs}

\addcontentsline{toc}{section}{References}
\bibliographystyle{yahapj}
\bibliography{main,lsst}

\clearpage

\printglossaries

\end{document}

%% file: aglossary.tex
\newacronym{AAAC} {AAAC} {Astronomy and Astrophysics Advisory Committee}
\newacronym{AAS} {AAS} {American Astronomical Society}
\newacronym{AEON} {AEON} {Alert Event Observatory Network}
\newacronym{ALMA} {ALMA} {Atacama Large Millimeter Array (\gls{ESO})}
\newacronym{API} {API} {Application Programming Interface}
\newacronym{APT} {APT} {Astronomer's Proposal Tool}
\newacronym{ASAS-SN} {ASAS-SN} {All-Sky Automated Survey for Supernovae}
\newacronym{ASTRON} {ASTRON} {Netherlands Institute for Radio Astronomy}
\newacronym{ATLAS} {ATLAS} {Asteroid Terrestrial-impact Last \gls{Alert} System}
\newacronym{AURA} {AURA} {\gls{Association of Universities for Research in Astronomy}}
\newacronym{AWS} {AWS} {Amazon Web Services}
\newglossaryentry{Alert} {name={Alert}, description={A packet of information for each source detected with signal-to-noise ratio > 5 in a difference image by Alert Production, containing measurement and characterization parameters based on the past 12 months of LSST observations plus small cutouts of the single-visit, template, and difference images, distributed via the internet}}
\newglossaryentry{Alert Production} {name={Alert Production}, description={Executing on the Prompt Processing system, the Alert Production payload processes and calibrates incoming images, performs Difference Image Analysis to identify DIASources and DIAObjects, and then packages the resulting alerts for distribution.}}
\newglossaryentry{Archive} {name={Archive}, description={The repository for documents required by the NSF to be kept. These include documents related to design and development, construction, integration, test, and operations of the LSST observatory system. The archive is maintained using the enterprise content management system DocuShare, which is accessible through a link on the project website www.project.lsst.org}}
\newglossaryentry{Archive Center} {name={Archive Center}, description={Part of the LSST Data Management System, the LSST archive center is a data center at NCSA that hosts the LSST Archive, which includes released science data and metadata, observatory and engineering data, and supporting software such as the LSST Software Stack}}
\newglossaryentry{Association Pipeline} {name={Association Pipeline}, description={An application that matches detected Sources or DIASources or generated Objects to an existing catalog of Objects, producing a (possibly many-to-many) set of associations and a list of unassociated inputs. Association Pipelines are used in Alert Production after DIASource generation and in the final stages of Data Release processing to ensure continuity of Object identifiers}}
\newglossaryentry{Association of Universities for Research in Astronomy} {name={Association of Universities for Research in Astronomy}, description={ consortium of US institutions and international affiliates that operates world-class astronomical observatories, AURA is the legal entity responsible for managing what it calls independent operating Centers, including LSST, under respective cooperative agreements with the National Science Foundation. AURA assumes fiducial responsibility for the funds provided through those cooperative agreements. AURA also is the legal owner of the AURA Observatory properties in Chile}}
\newacronym{Astro Data Lab} {Astro Data Lab} {Astro Data Lab science platform}
\newglossaryentry{Authentication} {name={Authentication}, description={The action of demonstrating who you are and an person, mission, or other entity. Usually by use of a password or security token}}
\newacronym{B} {B} {Byte (8 bit)}
\newacronym{BCE} {BCE} {Before Common Era}
\newglossaryentry{Butler} {name={Butler}, description={A middleware component for persisting and retrieving image datasets (raw or processed), calibration reference data, and catalogs}}
\newacronym{CA} {CA} {Control (or Cost) Account}
\newacronym{CANFAR} {CANFAR} {Canadian Advanced Network for Astronomical Research}
\newacronym{CARTA} {CARTA} {Cube Analysis and Rendering Tool for Astronomy}
\newacronym{CCD} {CCD} {\gls{Charge-Coupled Device}}
\newacronym{CL} {CL} {Command Language}
\newacronym{CMB} {CMB} {Cosmic Microwave Background}
\newacronym{CMB-S4} {CMB-S4} {Cosmic Microwave Background Stage 4}
\newacronym{CO} {CO} {Carbon Monoxide}
\newacronym{CPU} {CPU} {Central Processing Unit}
\newglossaryentry{Camera} {name={Camera}, description={The LSST subsystem responsible for the 3.2-gigapixel LSST camera, which will take more than 800 panoramic images of the sky every night. SLAC leads a consortium of Department of Energy laboratories to design and build the camera sensors, optics, electronics, cryostat, filters and filter exchange mechanism, and camera control system}}
\newglossaryentry{Center} {name={Center}, description={An entity managed by AURA that is responsible for execution of a federally funded project}}
\newglossaryentry{Charge-Coupled Device} {name={Charge-Coupled Device}, description={a particular kind of solid-state sensor for detecting optical-band photons. It is composed of a 2-D array of pixels, and one or more read-out amplifiers}}
\newglossaryentry{Commissioning} {name={Commissioning}, description={A two-year phase at the end of the Construction project during which a technical team a) integrates the various technical components of the three subsystems; b) shows their compliance with ICDs and system-level requirements as detailed in the LSST Observatory System Specifications document (OSS, LSE-30); and c) performs science verification to show compliance with the survey performance specifications as detailed in the LSST Science Requirements Document (SRD, LPM-17)}}
\newglossaryentry{Construction} {name={Construction}, description={The period during which LSST observatory facilities, components, hardware, and software are built, tested, integrated, and commissioned. Construction follows design and development and precedes operations. The LSST construction phase is funded through the NSF MREFC account}}
\newacronym{DB} {DB} {DataBase}
\newacronym{DC} {DC} {Data \gls{Center}}
\newacronym{DCR} {DCR} {\gls{Differential Chromatic Refraction}}
\newacronym{DES} {DES} {Dark Energy Survey}
\newacronym{DESI} {DESI} {Dark Energy Spectroscopic Instrument}
\newacronym{DIA} {DIA} {\gls{Difference Image Analysis}}
\newglossaryentry{DIAObject} {name={DIAObject}, description={A DIAObject is the association of DIASources, by coordinate, that have been detected with signal-to-noise ratio greater than 5 in at least one difference image. It is distinguished from a regular Object in that its brightness varies in time, and from a SSObject in that it is stationary (non-moving)}}
\newglossaryentry{DIASource} {name={DIASource}, description={A DIASource is a detection with signal-to-noise ratio greater than 5 in a difference image}}
\newacronym{DM} {DM} {\gls{Data Management}}
\newacronym{DMS} {DMS} {\gls{Data Management Subsystem}}
\newacronym{DOE} {DOE} {\gls{Department of Energy}}
\newacronym{DPAC} {DPAC} {Data Processing and Analysis Consortium (\gls{Gaia})}
\newacronym{DR} {DR} {\gls{Data Release}}
\newacronym{DRAGONS} {DRAGONS} {Data Reduction for Astronomy from Gemini Observatory North and South}
\newacronym{DRP} {DRP} {\gls{Data Release Production}}
\newglossaryentry{Data Management} {name={Data Management}, description={The LSST Subsystem responsible for the Data Management System (DMS), which will capture, store, catalog, and serve the LSST dataset to the scientific community and public. The DM team is responsible for the DMS architecture, applications, middleware, infrastructure, algorithms, and Observatory Network Design. DM is a distributed team working at LSST and partner institutions, with the DM Subsystem Manager located at LSST headquarters in Tucson}}
\newglossaryentry{Data Management Subsystem} {name={Data Management Subsystem}, description={The Data Management Subsystem is one of the four subsystems which constitute the LSST Construction Project. The Data Management Subsystem is responsible for developing and delivering the LSST Data Management System to the LSST Operations Project}}
\newglossaryentry{Data Management System} {name={Data Management System}, description={The computing infrastructure, middleware, and applications that process, store, and enable information extraction from the LSST dataset; the DMS will process peta-scale data volume, convert raw images into a faithful representation of the universe, and archive the results in a useful form. The infrastructure layer consists of the computing, storage, networking hardware, and system software. The middleware layer handles distributed processing, data access, user interface, and system operations services. The applications layer includes the data pipelines and the science data archives' products and services}}
\newglossaryentry{Data Release} {name={Data Release}, description={The approximately annual reprocessing of all LSST data, and the installation of the resulting data products in the LSST Data Access Centers, which marks the start of the two-year proprietary period}}
\newglossaryentry{Data Release Production} {name={Data Release Production}, description={An episode of (re)processing all of the accumulated LSST images, during which all output DR data products are generated. These episodes are planned to occur annually during the LSST survey, and the processing will be executed at the Archive Center. This includes Difference Imaging Analysis, generating deep Coadd Images, Source detection and association, creating Object and Solar System Object catalogs, and related metadata}}
\newglossaryentry{Department of Energy} {name={Department of Energy}, description={cabinet department of the United States federal government; the DOE has assumed technical and financial responsibility for providing the LSST camera. The DOE's responsibilities are executed by a collaboration led by SLAC National Accelerator Laboratory}}
\newglossaryentry{Difference Image} {name={Difference Image}, description={Refers to the result formed from the pixel-by-pixel difference of two images of the sky, after warping to the same pixel grid, scaling to the same photometric response, matching to the same PSF shape, and applying a correction for Differential Chromatic Refraction. The pixels in a difference thus formed should be zero (apart from noise) except for sources that are new, or have changed in brightness or position. In the LSST context, the difference is generally taken between a visit image and template. }}
\newglossaryentry{Difference Image Analysis} {name={Difference Image Analysis}, description={The detection and characterization of sources in the Difference Image that are above a configurable threshold, done as part of Alert Generation Pipeline}}
\newglossaryentry{Differential Chromatic Refraction} {name={Differential Chromatic Refraction}, description={The refraction of incident light by Earth's atmosphere causes the apparent position of objects to be shifted, and the size of this shift depends on both the wavelength of the source and its airmass at the time of observation. DCR corrections are done as a part of DIA}}
\newglossaryentry{Docker} {name={Docker}, description={A system for packaging and distributing software using self-contained containers which may be run on any Linux system; \url{https://www.docker.com/}}}
\newglossaryentry{DocuShare} {name={DocuShare}, description={The trade name for the enterprise management software used by LSST to archive and manage documents}}
\newglossaryentry{Document} {name={Document}, description={Any object (in any application supported by DocuShare or design archives such as PDMWorks or GIT) that supports project management or records milestones and deliverables of the LSST Project}}
\newacronym{ELTP} {ELTP} {Extremely Large Telescope Program}
\newacronym{EPICS} {EPICS} {Experimental Physics and Industrial Control System}
\newacronym{ESA} {ESA} {European Space Agency}
\newacronym{ESAP} {ESAP} {ESFRI Science Analysis Platform}
\newacronym{ESCAPE} {ESCAPE} {European Science Cluster of Astronomy \\& Particle physics \gls{ESFRI} research infrastructures}
\newacronym{ESFRI} {ESFRI} {European Strategy Forum on Research Infrastructures}
\newacronym{ESO} {ESO} {European Southern Observatory}
\newacronym{FADI} {FADI} {Future of Astronomical Data Infrastructure}
\newacronym{FAIR} {FAIR} {Findable, Accessible, Interoperable and Reusable}
\newacronym{FITS} {FITS} {\gls{Flexible Image Transport System}}
\newacronym{FTE} {FTE} {\gls{Full-Time Equivalent}}
\newglossaryentry{Firefly} {name={Firefly}, description={A framework of software components written by IPAC for building web-based user interfaces to astronomical archives, through which data may be searched and retrieved, and viewed as FITS images, catalogs, and/or plots. Firefly tools will be integrated into the Science Platform}}
\newglossaryentry{Flexible Image Transport System} {name={Flexible Image Transport System}, description={an international standard in astronomy for storing images, tables, and metadata in disk files. See the IAU FITS Standard for details}}
\newglossaryentry{Full-Time Equivalent} {name={Full-Time Equivalent}, description={A unit equivalent to one person working full time for one year with normal holidays, vacations, and sick time. No paid overtime is assumed}}
\newacronym{GCN} {GCN} {GRB Coordinates Network}
\newacronym{GCP} {GCP} {Google Cloud Platform}
\newacronym{GPFS} {GPFS} {General Parallel File System (now \gls{IBM} Spectrum Scale)}
\newacronym{GPP} {GPP} {Gemini Program Platform}
\newacronym{GPU} {GPU} {Graphics Processing Unit}
\newacronym{GRB} {GRB} {Gamma-Ray Burst}
\newglossaryentry{Gaia} {name={Gaia}, description={a space observatory of the European Space Agency, launched in 2013 and expected to operate until 2025. The spacecraft is designed for astrometry: measuring the positions, distances and motions of stars with unprecedented precision}}
\newglossaryentry{General Parallel File System} {name={General Parallel File System}, description={The bulk data storage provided through a POSIX filesystem interface at the LSST Data Facility. Refers specifically to IBM’s General Parallel File System; also known as IBM Spectrum Scale}}
\newacronym{HDU} {HDU} {Header Data Unit}
\newacronym{HEASARC} {HEASARC} {NASA's \gls{Archive} of Data on Energetic Phenomena}
\newacronym{HI} {HI} {Hydrogen iodide}
\newacronym{HPC} {HPC} {High Performance Computing}
\newacronym{HST} {HST} {Hubble Space Telescope}
\newacronym{HTC} {HTC} {High Throughput Computing}
\newglossaryentry{Handle} {name={Handle}, description={The unique identifier assigned to a document uploaded to DocuShare}}
\newacronym{IAU} {IAU} {International Astronomical Union}
\newacronym{IBM} {IBM} {International Business Machines}
\newacronym{IPAC} {IPAC} {No longer an acronym; science and data center at Caltech}
\newacronym{IRAF} {IRAF} {\gls{Image Reduction and Analysis Facility}}
\newacronym{IRSA} {IRSA} {Infrared Science \gls{Archive} (NASA)}
\newacronym{IT} {IT} {Information Technology}
\newacronym{IVOA} {IVOA} {International Virtual-Observatory Alliance}
\newglossaryentry{Image Reduction and Analysis Facility} {name={Image Reduction and Analysis Facility}, description={  a collection of software written at the National Optical Astronomy Observatory (now NOIRLab) geared towards the reduction of astronomical images in pixel array form.}}
\newglossaryentry{J2000} {name={J2000}, description={Julian Date referring to the instant of 12 noon (midday) on January 1, 2000. IAU standard equinox.}}
\newacronym{JD} {JD} {\gls{Julian Date}}
\newglossaryentry{JIRA} {name={JIRA}, description={issue tracking product (not an acronym but a truncation of Gojira the Japanese name for Godzilla)}}
\newacronym{JWST} {JWST} {James Webb Space Telescope (formerly known as NGST)}
\newglossaryentry{Julian Date} {name={Julian Date}, description={The Julian Date (JD) of any instant is the Julian day number for the preceding noon (UTC), plus the fraction of the day elapsed since that instant. The Julian day number is a running sequence of integral days, starting at noon, since the beginning of the Julian Period; JD 0.0 corresponds to noon on 1 January 4713 BCE. Various Julian Date converters are available on the Web. For example, 18h 00m 00.0s UT on 2014-July-01 (near the start of LSST construction) corresponds to JD 2456840.25}}
\newglossaryentry{Kubernetes} {name={Kubernetes}, description={A system for automating application deployment and management using software containers (e.g. Docker); \url{https://kubernetes.io}}}
\newacronym{L0} {L0} {Level 0}
\newacronym{L1} {L1} {Level 1}
\newacronym{LDM} {LDM} {LSST Data Management (Document \gls{Handle})}
\newacronym{LINCC} {LINCC} {LSST Interdisciplinary Network for Collaboration and Computing}
\newacronym{LPM} {LPM} {LSST Project Management (Document \gls{Handle})}
\newacronym{LSE} {LSE} {LSST \gls{Systems Engineering} (Document Handle)}
\newacronym{LSST} {LSST} {Legacy Survey of Space and Time (formerly Large Synoptic Survey Telescope)}
\newglossaryentry{LSST Project Office} {name={LSST Project Office}, description={Official name of the stand-alone AURA operating center responsible for execution of the LSST construction project under the NSF MREFC account}}
\newglossaryentry{LSST Science Pipelines} {name={LSST Science Pipelines}, description={software used to perform the LSST data reduction pipelines.lsst.io}}
\newacronym{LSSTPO} {LSSTPO} {\gls{LSST Project Office}}
\newacronym{MAST} {MAST} {Mikulski \gls{Archive} for Space Telescopes}
\newacronym{MMA} {MMA} {Multi Messenger Astronomy}
\newacronym{MMT} {MMT} {Multiple Mirror Telescope}
\newacronym{MREFC} {MREFC} {\gls{Major Research Equipment and Facility Construction}}
\newglossaryentry{Major Research Equipment and Facility Construction} {name={Major Research Equipment and Facility Construction}, description={the NSF account through which large facilities construction projects such as LSST are funded}}
\newglossaryentry{Mapper} {name={Mapper}, description={A piece of software that abstracts persisting and unpersisting data; specifically, it knows how to navigate a data repository to locate data that match selection criteria that are relevant for data obtained with a particular camera. Used by the Butler}}
\newacronym{NASA} {NASA} {National Aeronautics and Space Administration}
\newacronym{NAVO} {NAVO} {NASA Astronomical Virtual Observatories}
\newacronym{NCSA} {NCSA} {National \gls{Center} for Supercomputing Applications}
\newacronym{NFS} {NFS} {Network File System}
\newacronym{NOIRLab} {NOIRLab} {NSF's National Optical-Infrared Astronomy Research Laboratory; \url{https://noirlab.edu}}
\newacronym{NSF} {NSF} {\gls{National Science Foundation}}
\newglossaryentry{National Science Foundation} {name={National Science Foundation}, description={primary federal agency supporting research in all fields of fundamental science and engineering; NSF selects and funds projects through competitive, merit-based review}}
\newacronym{OCS} {OCS} {Observatory Control System}
\newacronym{OSS} {OSS} {Observatory System Specifications; \gls{LSE}-30}
\newglossaryentry{Object} {name={Object}, description={In LSST nomenclature this refers to an astronomical object, such as a star, galaxy, or other physical entity. E.g., comets, asteroids are also Objects but typically called a Moving Object or a Solar System Object (SSObject). One of the DRP data products is a table of Objects detected by LSST which can be static, or change brightness or position with time}}
\newglossaryentry{Offer} {name={Offer}, description={A response to a solicitation that, if accepted, would bind the offeror to perform the work described in resultant contract. Responses to sealed bidding are offers that are often referred to as 'bids' or 'sealed bids;' responses to a request for proposals (RFP, negotiated-type procurements) are offers often referred to as 'proposals' responses to a request for quotations (RFQ) are not offers and are generally called 'quotes'}}
\newglossaryentry{Operations} {name={Operations}, description={The 10-year period following construction and commissioning during which the LSST Observatory conducts its survey}}
\newacronym{PD} {PD} {Program Development}
\newacronym{PDF} {PDF} {Probability Density Function}
\newacronym{PI} {PI} {Principle Investigator}
\newacronym{POSIX} {POSIX} {Portable Operating System Interface}
\newacronym{PSF} {PSF} {Point Spread Function}
\newglossaryentry{Project Manager} {name={Project Manager}, description={The person responsible for exercising leadership and oversight over the entire Rubin project; he or she controls schedule, budget, and all contingency funds}}
\newglossaryentry{Prompt Processing} {name={Prompt Processing}, description={The data processing which occurs at the Archive Center based on the stream of images coming from the telescope. This includes both Alert Production, which scans the image stream to identify and send alerts on transient and variable sources, and Solar System Processing, which identifies and characterizes objects in our solar system. It also includes specialized rapid calibration and Commissioning processing. Prompt Processing generates the Prompt Data Products.}}
\newacronym{Q3C} {Q3C} {Quad Tree Cube}
\newacronym{QA} {QA} {\gls{Quality Assurance}}
\newacronym{QC} {QC} {\gls{Quality Control}}
\newglossaryentry{Qserv} {name={Qserv}, description={LSST's distributed parallel database. This database system is used for collecting, storing, and serving LSST Data Release Catalogs and Project metadata, and is part of the Software Stack}}
\newglossaryentry{Quality Assurance} {name={Quality Assurance}, description={All activities, deliverables, services, documents, procedures or artifacts which are designed to ensure the quality of DM deliverables. This may include QC systems, in so far as they are covered in the charge described in LDM-622. Note that contrasts with the LDM-522 definition of “QA” as “Quality Analysis”, a manual process which occurs only during commissioning and operations. See also: Quality Control}}
\newglossaryentry{Quality Control} {name={Quality Control}, description={Services and processes which are aimed at measuring and monitoring a system to verify and characterize its performance (as in LDM-522). Quality Control systems run autonomously, only notifying people when an anomaly has been detected. See also Quality Assurance}}
\newacronym{RA} {RA} {Right Ascension}
\newacronym{RFP} {RFP} {Request For Proposals}
\newacronym{RFQ} {RFQ} {Request For Quotations}
\newglossaryentry{RSP} {name={RSP}, description={Rubin Science Platform}}
\newglossaryentry{Requirement} {name={Requirement}, description={A declaration of a specified function or quantitative performance that the delivered system or subsystem must meet.  It is a statement that identifies a necessary attribute, capability, characteristic, or quality of a system in order for the delivered system or subsystem to meet a derived or higher requirement, constraint, or function}}
\newglossaryentry{Rucio} {name={Rucio}, description={Rucio is a project that provides services and associated libraries for allowing scientific collaborations to manage large volumes of data spread across facilities at multiple institutions and organizations. Rucio has been developed by the ATLAS experiment}}
\newacronym{S3} {S3} {(Amazon) Simple Storage Service}
\newacronym{SDSS} {SDSS} {\gls{Sloan Digital Sky Survey}}
\newacronym{SIA} {SIA} {Simple Image Access (\gls{IVOA} standard)}
\newacronym{SLAC} {SLAC} {\gls{SLAC National Accelerator Laboratory}}
\newglossaryentry{SLAC National Accelerator Laboratory} {name={SLAC National Accelerator Laboratory}, description={A national laboratory funded by the US Department of Energy (DOE); SLAC leads a consortium of DOE laboratories that has assumed responsibility for providing the LSST camera. Although the Camera project manages its own schedule and budget, including contingency, the Camera team’s schedule and requirements are integrated with the larger Project.  The camera effort is accountable to the LSSTPO.}}
\newacronym{SN} {SN} {SuperNovae}
\newacronym{SP} {SP} {\gls{Science Platform}}
\newacronym{SQL} {SQL} {Structured Query Language}
\newacronym{SRD} {SRD} {LSST Science Requirements; \gls{LPM}-17}
\newacronym{SW} {SW} {Software (also denoted S/W)}
\newglossaryentry{Science Platform} {name={Science Platform}, description={A set of integrated web applications and services deployed at the LSST Data Access Centers (DACs) through which the scientific community will access, visualize, and perform next-to-the-data analysis of the LSST data products}}
\newglossaryentry{Scope} {name={Scope}, description={The work needed to be accomplished in order to deliver the product, service, or result with the specified features and functions}}
\newglossaryentry{Sloan Digital Sky Survey} {name={Sloan Digital Sky Survey}, description={is a digital survey of roughly 10,000 square degrees of sky around the north Galactic pole, plus a ~300 square degree stripe along the celestial equator}}
\newglossaryentry{Software Stack} {name={Software Stack}, description={Often referred to as the LSST Stack, or just The Stack, it is the collection of software written by the LSST Data Management Team to process, generate, and serve LSST images, transient alerts, and catalogs. The Stack includes the LSST Science Pipelines, as well as packages upon which the DM software depends. It is open source and publicly available}}
\newglossaryentry{Solar System Object} {name={Solar System Object}, description={A solar system object is an astrophysical object that is identified as part of the Solar System: planets and their satellites, asteroids, comets, etc. This class of object had historically been referred to within the LSST Project as Moving Objects}}
\newglossaryentry{Solar System Processing} {name={Solar System Processing}, description={A component of the Prompt Processing system, Solar System Processing identifies new SSObjects using unassociated DIASources.}}
\newglossaryentry{Source} {name={Source}, description={A single detection of an astrophysical object in an image, the characteristics for which are stored in the Source Catalog of the DRP database. The association of Sources that are non-moving lead to Objects; the association of moving Sources leads to Solar System Objects. (Note that in non-LSST usage "source" is often used for what LSST calls an Object.)}}
\newglossaryentry{Subsystem} {name={Subsystem}, description={A set of elements comprising a system within the larger LSST system that is responsible for a key technical deliverable of the project}}
\newglossaryentry{Subsystem Manager} {name={Subsystem Manager}, description={responsible manager for an LSST subsystem; he or she exercises authority, within prescribed limits and under scrutiny of the Project Manager, over the relevant subsystem's cost, schedule, and work plans}}
\newglossaryentry{Systems Engineering} {name={Systems Engineering}, description={an interdisciplinary field of engineering that focuses on how to design and manage complex engineering systems over their life cycles. Issues such as requirements engineering, reliability, logistics, coordination of different teams, testing and evaluation, maintainability and many other disciplines necessary for successful system development, design, implementation, and ultimate decommission become more difficult when dealing with large or complex projects. Systems engineering deals with work-processes, optimization methods, and risk management tools in such projects. It overlaps technical and human-centered disciplines such as industrial engineering, control engineering, software engineering, organizational studies, and project management. Systems engineering ensures that all likely aspects of a project or system are considered, and integrated into a whole}}
\newacronym{TAC} {TAC} {Time Allocation Committee}
\newacronym{TAP} {TAP} {Table Access Protocol (\gls{IVOA} standard)}
\newacronym{TDA} {TDA} {Time Domain Astronomy}
\newacronym{TMT} {TMT} {Thirty Meter Telescope}
\newacronym{TOM} {TOM} {Target and Observation Manager}
\newacronym{UI} {UI} {User Interface}
\newacronym{UK} {UK} {United Kingdom}
\newacronym{US} {US} {United States}
\newacronym{UT} {UT} {Universal Time}
\newacronym{UTC} {UTC} {Coordinated Universal Time}
\newacronym{VLT} {VLT} {Very Large Telescope (\gls{ESO})}
\newacronym{VO} {VO} {Virtual Observatory}
\newacronym{WCS} {WCS} {\gls{World Coordinate System}}
\newglossaryentry{World Coordinate System} {name={World Coordinate System}, description={a mapping from image pixel coordinates to physical coordinates; in the case of images the mapping is to sky coordinates, generally in an equatorial (RA, Dec) system. The WCS is expressed in FITS file extensions as a collection of header keyword=value pairs (basically, the values of parameters for a selected functional representation of the mapping) that are specified in the FITS Standard}}
\newacronym{XRAC} {XRAC} {XSEDE Resource Allocation Committee}
\newacronym{XSEDE} {XSEDE} {Extreme Science and Engineering Discovery Environment}
\newglossaryentry{airmass} {name={airmass}, description={The pathlength of light from an astrophysical source through the Earth's atmosphere. It is given approximately by sec z, where z is the angular distance from the zenith (the point directly overhead, where airmass = 1.0) to the source}}
\newglossaryentry{astrometry} {name={astrometry}, description={In astronomy, the sub-discipline of astrometry concerns precision measurement of positions (at a reference epoch), and real and apparent motions of astrophysical objects. Real motion means 3-D motions of the object with respect to an inertial reference frame; apparent motions are an artifact of the motion of the Earth. Astrometry per se is sometimes confused with the act of determining a World Coordinate System (WCS), which is a functional characterization of the mapping from pixels in an image or spectrum to world coordinate such as (RA, Dec) or wavelength}}
\newglossaryentry{astronomical object} {name={astronomical object}, description={A star, galaxy, asteroid, or other physical object of astronomical interest. Beware: in non-LSST usage, these are often known as sources}}
\newglossaryentry{calibration} {name={calibration}, description={The process of translating signals produced by a measuring instrument such as a telescope and camera into physical units such as flux, which are used for scientific analysis. Calibration removes most of the contributions to the signal from environmental and instrumental factors, such that only the astronomical component remains}}
\newglossaryentry{camera} {name={camera}, description={An imaging device mounted at a telescope focal plane, composed of optics, a shutter, a set of filters, and one or more sensors arranged in a focal plane array}}
\newglossaryentry{cloud} {name={cloud}, description={A visible mass of condensed water vapor floating in the atmosphere, typically high above the ground or in interstellar space acting as the birthplace for stars.  Also a way of computing (on other peoples computers leveraging their services and availability).}}
\newglossaryentry{community software} {name={community software}, description={Software developed for and shared among a large group of relatively like-minded users (e.g. astronomers). Typically, but not necessarily, open source software and open development-based.}}
\newglossaryentry{configuration} {name={configuration}, description={A task-specific set of configuration parameters, also called a 'config'. The config is read-only; once a task is constructed, the same configuration will be used to process all data. This makes the data processing more predictable: it does not depend on the order in which items of data are processed. This is distinct from arguments or options, which are allowed to vary from one task invocation to the next}}
\newglossaryentry{data collection} {name={data collection}, description={A data collection in the second-generation (Gen2) Butler (referred to as a data repository in earlier generations) consists of hierarchically organized data files, an inventory or registry of the contents (i.e., metadata from the data files) stored in an sqlite3 file, and a Mapper file that specifies to the LSST Stack software the camera model to apply when accessing the data in the data repository}}
\newglossaryentry{data repository} {name={data repository}, description={A data repository consists of hierarchically organized data files, an inventory or registry of the contents (i.e., metadata from the data files) stored in an sqlite3 file, and a Mapper file that specifies to the LSST Stack software the camera model to apply when accessing the data in the repository. With the second-generation (Gen2) Butler, the term repository will be replaced by data collection}}
\newglossaryentry{ephemeris} {name={ephemeris}, description={An ephemeris (pl: ephemerides) gives the predicted positions of astronomical objects or artificial satellites in the sky with time. The ephemerides are computed from mathematical models of motion of the object and the Earth. In LSST Solar System Processing, it refers to a predicted position (RA/Dec/time/etc) of a Solar System Object (SSObject)}}
\newglossaryentry{epoch} {name={epoch}, description={Sky coordinate reference frame, e.g., J2000. Alternatively refers to a single observation (usually photometric, can be multi-band) of a variable source}}
\newglossaryentry{flux} {name={flux}, description={Shorthand for radiative flux, it is a measure of the transport of radiant energy per unit area per unit time. In astronomy this is usually expressed in cgs units: erg/cm2/s}}
\newglossaryentry{git} {name={git}, description={A distributed revision control system, often used for software source code. See the Git User Manual for details. Not developed by LSST DM}}
\newglossaryentry{interoperability} {name={interoperability}, description={the ability of systems or software to exchange and make use of information between them.}}
\newglossaryentry{metadata} {name={metadata}, description={General term for data about data, e.g., attributes of astronomical objects (e.g. images, sources, astroObjects, etc.) that are characteristics of the objects themselves, and facilitate the organization, preservation, and query of data sets. (E.g., a FITS header contains metadata)}}
\newglossaryentry{middleware} {name={middleware}, description={Software that acts as a bridge between other systems or software usually a database or network. Specifically in the Data Management System this refers to Butler for data access and Workflow management for distributed processing.}}
\newglossaryentry{monitoring} {name={monitoring}, description={In DM QA, this refers to the process of collecting, storing, aggregating and visualizing metrics}}
\newglossaryentry{open development} {name={open development}, description={A process for developing software that emphasizes all code contribution and decision-making be done in the open, available to as wide a group as possible (This usually means anyone with internet access).}}
\newglossaryentry{open source software} {name={open source software}, description={Open source software is a type of software in which source code is released under a license in which the copyright holder grants users the rights to study, change, and distribute the software to anyone and for any purpose. Note that this is \emph{not} necessarily the same as open to contribution (see open development).}}
\newglossaryentry{pipeline} {name={pipeline}, description={A configured sequence of software tasks (Stages) to process data and generate data products. Example: Association Pipeline}}
\newglossaryentry{shape} {name={shape}, description={In reference to a Source or Object, the shape is a functional characterization of its spatial intensity distribution, and the integral of the shape is the flux. Shape characterizations are a data product in the DIASource, DIAObject, Source, and Object catalogs}}
\newglossaryentry{software} {name={software}, description={The programs and other operating information used by a computer.}}
\newglossaryentry{sqlite3} {name={sqlite3}, description={A software package external to DM, sqlite3 provides a SQL interface compliant with the DB-API 2.0 specification for SQLite, a self-contained public-domain SQL database engine}}
\newglossaryentry{stack} {name={stack}, description={a grouping, usually in layers (hence stack), of software packages and services to achieve a common goal. Often providing a higher level set of end user oriented services and tools}}
\newglossaryentry{transient} {name={transient}, description={A transient source is one that has been detected on a difference image, but has not been associated with either an astronomical object or a solar system body}}

%% file: authors.tex

\author{~Michael~R.~Blanton}
\affiliation{Center for Cosmology \& Particle Physics, New York University, 726 Broadway, New York, 10003, USA}

\author[0000-0003-3509-0870]{~Janet~D.~Evans}
\affiliation{Center for Astrophysics, Harvard \& Smithsonian, 60 Garden Street, Cambridge, MA 02138}

\author[0000-0001-8452-9574]{~Dara~Norman}
\affiliation{NSF's National Optical-Infrared Astronomy Research Laboratory, 950 N.\ Cherry Ave., Tucson, AZ 85719, USA}

\author[0000-0003-4141-6195]{~William~O'Mullane}
\affiliation{Rubin Observatory Project Office, 950 N.\ Cherry Ave., Tucson, AZ  85719, USA}

\author[0000-0003-0872-7098]{~Adrian~Price-Whelan}
\affiliation{Center for Computational Astrophysics, Flatiron Institute, 162 Fifth Ave, New York, NY, 10010, USA}

\author{~Luca~Rizzi}
\affiliation{National Science Foundation, 2415 Eisenhower Avenue, Alexandria, Virginia 22314, USA}

\author{~Alberto~Accomazzi}
\affiliation{Center for Astrophysics, Harvard \& Smithsonian, 60 Garden Street, Cambridge, MA 02138}

\author{~Megan~Ansdell}
\affiliation{National Aeronautics and Space Administration, Goddard Space Flight Center, Greenbelt, Maryland, USA}

\author{~Stephen~Bailey}
\affiliation{Lawrence Berkeley National Laboratory, 1 Cyclotron Road, Berkeley, CA 94720, USA}

\author{~Paul~Barrett}
\affiliation{US Naval Observatory Flagstaff Station, 10391 Naval Observatory Road, Flagstaff, AZ 86001, USA}

\author{~Steven~Berukoff}
\affiliation{NSF's National Optical-Infrared Astronomy Research Laboratory, 950 N.\ Cherry Ave., Tucson, AZ 85719, USA}

\author{~Adam~Bolton}
\affiliation{NSF's National Optical-Infrared Astronomy Research Laboratory, 950 N.\ Cherry Ave., Tucson, AZ 85719, USA}

\author{~Julian~Borrill}
\affiliation{Lawrence Berkeley National Laboratory, 1 Cyclotron Road, Berkeley, CA 94720, USA}

\author{~Kelle~Cruz}
\affiliation{City University of New York,  Hunter College, Hunter North Building, 695 Park Ave., New York, NY 10065, USA}

\author{~Julianne~Dalcanton}
\affiliation{Center for Computational Astrophysics, Flatiron Institute, 162 Fifth Ave, New York, NY, 10010, USA}

\author{~Vandana~Desai}
\affiliation{IPAC, California Institute of Technology, MS 100-22, Pasadena, CA 91125, USA}

\author[0000-0003-1598-6979]{~Gregory~P.~Dubois-Felsmann}
\affiliation{IPAC, California Institute of Technology, MS 100-22, Pasadena, CA 91125, USA}

\author[0000-0002-8333-7615]{~Frossie~Economou}
\affiliation{Rubin Observatory Project Office, 950 N.\ Cherry Ave., Tucson, AZ  85719, USA}

\author[0000-0001-7113-2738]{~Henry~Ferguson}
\affiliation{Space Telescope Science Institute, 3700 San Martin Drive, Baltimore, MD 21218, USA}

\author{~Bryan~Field}
\affiliation{Office of Science, U.S. Department of Energy, 1000 Independence Ave., SW Washington, DC 20585, USA}

\author{~Dan~Foreman-Mackey}
\affiliation{Center for Computational Astrophysics, Flatiron Institute, 162 Fifth Ave, New York, NY, 10010, USA}

\author{~Jaime~Forero-Romero}
\affiliation{Universidad de los Andes, Cra 1 Num. 18A - 12 Bogot\'{a} - Colombia}

\author{~Niall~Gaffney}
\affiliation{Texas Advanced Computing Center, The University of Texas at Austin, USA.}

\author{~Kim~Gillies}
\affiliation{TMT International Observatory}

\author[0000-0002-3168-0139]{~Matthew~J.~Graham}
\affiliation{Astronomy Department, California Institute of Technology, 1200 East California Blvd., Pasadena CA 91125, USA}

\author{~Steven~Gwyn}
\affiliation{National Research Council Canada, 5071 West Saanich Road, Victoria, Canada}

\author{~Joseph~Hennawi}
\affiliation{University of California, Santa Barbara, CA 93106-9530, USA}

\author{~Anna~L.~H.~Hughes}
\affiliation{National Solar Observatory, 3665 Discovery Drive, Boulder, CO 80303, USA}

\author{~Tess~Jaffe}
\affiliation{National Aeronautics and Space Administration, Goddard Space Flight Center, Greenbelt, Maryland, USA}

\author{~Preshanth~Jagannathan}
\affiliation{National Radio Astronomy Observstory, 520 Edgemont Road, Charlottesville, VA 22903, USA}

\author[0000-0001-5982-167X]{~Tim~Jenness}
\affiliation{Rubin Observatory Project Office, 950 N.\ Cherry Ave., Tucson, AZ  85719, USA}

\author[0000-0003-1996-9252]{~Mario~Juri\'{c}}
\affiliation{Department of Astronomy and the DIRAC Institute, University of Washington, 3910 15th Avenue NE, Seattle, WA 98195, USA}

\author{~JJ~Kavelaars}
\affiliation{Canadian Astronomy Data Centre, 5071 West Saanich Road, Victoria, British Columbia V9E 2E7, Canada}

\author{~Kerk~Kee}
\affiliation{College of Media and Communication, Texas Tech University, Lubbock, TX 79409, USA}

\author[0000-0003-3221-0419]{~Jeff~Kern}
\affiliation{National Radio Astronomy Observstory, 520 Edgemont Road, Charlottesville, VA 22903, USA}

\author{~Anthony~Kremin}
\affiliation{Lawrence Berkeley National Laboratory, 1 Cyclotron Road, Berkeley, CA 94720, USA}

\author{~Kathleen~Labrie}
\affiliation{NSF's National Optical-Infrared Astronomy Research Laboratory, 950 N.\ Cherry Ave., Tucson, AZ 85719, USA}

\author{~Mark~Lacy}
\affiliation{National Radio Astronomy Observstory, 520 Edgemont Road, Charlottesville, VA 22903, USA}

\author{~Casey~Law}
\affiliation{Astronomy Department, California Institute of Technology, 1200 East California Blvd., Pasadena CA 91125, USA}

\author{~Rafael~Martínez-Galarza}
\affiliation{Department of Astronomy, Center for Astrophysics, Harvard University, 60 Garden St., Cambridge, MA 02138, USA}

\author{~Curtis~McCully}
\affiliation{Las Cumbres Observatory, 6740 Cortona Dr., Suite 102, Goleta, CA 93117, USA}

\author{~Julie~McEnery}
\affiliation{National Aeronautics and Space Administration, Goddard Space Flight Center, Greenbelt, Maryland, USA}

\author{~Bryan~Miller}
\affiliation{NSF's National Optical-Infrared Astronomy Research Laboratory, 950 N.\ Cherry Ave., Tucson, AZ 85719, USA}

\author{~Christopher~Moriarty}
\affiliation{Center for Astrophysics, Harvard \& Smithsonian, 60 Garden Street, Cambridge, MA 02138}

\author{~August~Muench}
\affiliation{American Astronomical Society}

\author{~Demitri~Muna}
\affiliation{National Aeronautics and Space Administration, Goddard Space Flight Center, Greenbelt, Maryland, USA}

\author{~Angela~Murillo}
\affiliation{IUPUI, 535 W. Michigan Street, IT 400, Indianapolis, IN, USA}

\author{~Gautham~Narayan}
\affiliation{University of Illinois, Physics and Astronomy Departments, 1110 W.\ Green St., Urbana, IL  61801, USA}

\author{~James~D.~Neill}
\affiliation{Division of Physics, Mathematics and Astronomy, California Institute of Technology, Pasadena, CA 91125, USA}

\author[0000-0002-7052-6900]{~Robert~Nikutta}
\affiliation{NSF's National Optical-Infrared Astronomy Research Laboratory, 950 N.\ Cherry Ave., Tucson, AZ 85719, USA}

\author{~Roopesh~Ojha}
\affiliation{National Aeronautics and Space Administration, Goddard Space Flight Center, Greenbelt, Maryland, USA}

\author[0000-0002-7134-8296]{~Knut~Olsen}
\affiliation{NSF's National Optical-Infrared Astronomy Research Laboratory, 950 N.\ Cherry Ave., Tucson, AZ 85719, USA}

\author{~John~O'Meara}
\affiliation{W. M. Keck Observatory, 65-1120 Mamalahoa Hwy.  Kamuela, HI 96743, USA}

\author{~Ben~Rusholme}
\affiliation{IPAC, California Institute of Technology, MS 100-22, Pasadena, CA 91125, USA}

\author{~Robert~Seaman}
\affiliation{University of Arizona, Tucson, AZ 85721, USA}

\author{~Nathaniel~Starkman}
\affiliation{University of Toronto, 27 King's College Circle, Toronto, Ontario M5S 1A1 Canada}

\author{~Martin~Still}
\affiliation{National Science Foundation, 2415 Eisenhower Avenue, Alexandria, Virginia 22314, USA}

\author{~Felix~Stoehr}
\affiliation{European Southern Observatory, Karl-Schwarzschild-Strasse 2, 85748 Garching bei München, Germany}

\author[0000-0001-9445-1846]{~John~D.~Swinbank}
\affiliation{ASTRON, Oude Hoogeveensedijk 4, 7991\,PD, Dwingeloo, The Netherlands}
\affiliation{Department of Astrophysical Sciences, Princeton University, Princeton, NJ 08544, USA}

\author{~Peter~Teuben}
\affiliation{University of Maryland, College Park, MD 20742}

\author{~Ignacio~Toledo}
\affiliation{Atacama Large Millimeter/submillimeter Array, San Pedro de Atacama, Antofagasta, Chile}

\author{~Erik~Tollerud}
\affiliation{Space Telescope Science Institute, 3700 San Martin Drive, Baltimore, MD 21218, USA}

\author{~Matthew~D.~Turk}
\affiliation{University of Illinois, Physics and Astronomy Departments, 1110 W.\ Green St., Urbana, IL  61801, USA}

\author{~James~Turner}
\affiliation{NSF's National Optical-Infrared Astronomy Research Laboratory, 950 N.\ Cherry Ave., Tucson, AZ 85719, USA}

\author{~William~Vacca}
\affiliation{NSF's National Optical-Infrared Astronomy Research Laboratory, 950 N.\ Cherry Ave., Tucson, AZ 85719, USA}

\author{~Joaquin~Vieira}
\affiliation{University of Illinois, Department of Astronomy, 1110 W.\ Green St., Urbana, IL  61801, USA}

\author{~Benjamin~Weaver}
\affiliation{NSF's National Optical-Infrared Astronomy Research Laboratory, 950 N.\ Cherry Ave., Tucson, AZ 85719, USA}

\author{~Benjamin~Weiner}
\affiliation{MMT Observatory, 1540 E. Second Street, University of Arizona, Tucson, AZ 85721-0064, USA}

\author{~Jason~Weiss}
\affiliation{TMT International Observatory}

\author{~Kyle~Westfall}
\affiliation{UC Observatories, UC Santa Cruz, 1156 High Street, Santa Cruz, CA 95064}

\author[0000-0003-2892-9906]{~Beth~Willman}
\affiliation{Rubin Observatory Project Office, 950 N.\ Cherry Ave., Tucson, AZ  85719, USA}
\affiliation{Steward Observatory, The University of Arizona, 933 N.\ Cherry Ave., Tucson, AZ 85721, USA}

\author{~Lily~Zhao}
\affiliation{Center for Computational Astrophysics, Flatiron Institute, 162 Fifth Ave, New York, NY, 10010, USA}

%% file: execsum.tex
\section{Executive Summary} \label{sec:execsum}

The astronomical community is grappling with the increasing volume and complexity of data produced by modern telescopes, due to difficulties in reducing, accessing, analyzing, and combining archives of data. To address this challenge, we propose the establishment of a coordinating body, an ``entity,'' with the specific mission of enhancing the interoperability, archiving, distribution, and production of both astronomical data and software.

This report is the culmination of a workshop held in February 2023 on the Future of Astronomical Data Infrastructure. Attended by 70 scientists and software professionals from ground-based and space-based missions and archives spanning the entire spectrum of astronomical research, the group deliberated on the prevailing state of software and data infrastructure in astronomy, identified pressing issues, and explored potential solutions.

In this report, we describe the ecosystem of astronomical data, its existing flaws, and the many gaps, duplication, inconsistencies, barriers to access, drags on productivity, missed opportunities, and risks to the long-term integrity of essential data sets. We also highlight  the successes and failures in a set of deep dives into several different illustrative components of the ecosystem, included as an appendix.

The report also puts forth a roadmap for the development of the proposed coordination entity, offering four potential structural models for consideration. Alongside, we present preliminary case studies scrutinizing their associated costs, advantages, disadvantages, and risks. We advocate for the establishment of a Steering Committee, bolstered by support from federal funding agencies, to take the lead in defining the entity's scope, requirements, and structural framework. This committee should actively engage with and solicit feedback from the broader astronomical community and the expert cadre specializing in astronomical data, culminating in the formulation of an actionable implementation plan for the relevant agencies.

The report emphasizes that the proposed entity should have broad community buy-in and the support of existing astronomical projects, missions, and public-facing archives, and not be perceived to be in competition with these existing organizations. The problems laid out in this report pose no fundamental technical show-stoppers, but solving these problems remains outside the funded scope of work of any existing organization. We urge the funding agencies to create an entity with the mission to address the system’s current flaws and with the resources to do so, unlocking enormous potential for new scientific discovery in this extraordinarily data-rich era of astronomical science.

%% file: intro.tex
\section{Introduction} \label{sec:intro}


\Lead{Frossie Economou}
\Contributors{Paul Barrett, Roopesh Ojha, Gautham Narayan, Demitri Muna}

\subsection{Overall Context and Audience(s)}


The final report of the Decadal Survey on Astronomy and Astrophysics 2020 \citep[Astro2020]{NAP26141}, published in November 2021, emphasizes the foundational role that data plays in modern astronomy. It highlights the urgent need to be prepared for the ``second wave of the data revolution,'' driven by the increasing number of survey facilities coming online. In Section 4.5, Astro2020 points out that ``progress will come from an end-to-end approach that considers the entire flow of data from the instrument, to the archive, to analysis and publication.'' It offers two recommendations:

\begin{itemize}
    \item NASA and the National Science Foundation should explore mechanisms to improve coordination among U.S. archive centers and to create a centralized nexus for interacting with the international archive communities. The goals of this effort should be informed by the broad scientific needs of the astronomical community.
    \item The National Science Foundation and stakeholders should develop a plan to address how to design, build, deploy, and sustain pipelines for producing science- ready data across all general-purpose ground-based observatories (both federally and privately funded), providing funding in exchange for ensuring that all pipelined observations are archived in a standard format for eventual public use.
\end{itemize}

In response to these recommendations, the Center for Computational Astrophysics at the Flatiron Institute, in collaboration with the \gls{NSF}, organized a workshop in February 2023 in New York City. The workshop aimed to identify technical solutions, implementation options, a long-term vision, and a tentative roadmap.

This document summarizes the outcomes of the workshop and has two primary intended audiences:
\begin{enumerate}
    \item Actors who have the remit and the agency to pursue the recommendations of this report. These may include current advisory bodies such as the AAAC, the relevant federal funding agencies (\gls{NSF}, \gls{NASA}, \gls{DOE}) supporting astronomical funding, and private consortia practicing in the field.
    \item Future bodies or community initiatives focused on pipelines and archives or other aspects of the software ecosystem that might take this report as guidance for their activities. These may include the coordinating entity envisaged in this report, or other activities that are conceived in support of the same goal in the future.
\end{enumerate}

The primary conclusion of the workshop is that the community that represents data and software infrastructure has a strong desire to work collegially to develop the elements of a shared ecosystem, but lacks the means to coordinate the effort and the appropriate financial support, which is not only insufficient, but also focused on supporting individual missions rather than shared solutions. 

Chapter \ref{sec:ecosys} provides a description of the main components of the astronomical data and software ecosystem, focusing on the concept of the ``cycle of discovery'' for individual investigators and extending it to other scientific cases such as surveys.
Chapter \ref{sec:problem} describes the issues that have affected the development of an effective data and software ecosystem. In response to these issues and to the recommendations of Astro2020, Chapter \ref{sec:scope} presents a possible roadmap to address these issue, and suggests the creation of a new ``coordination entity'', the nature of which should be determined by the work of a Steering Committee. As a starting point, Chapter \ref{sec:casestudies} (and the following 4 Chapters) describe 4 possible implementations of such an entity.


%% file: problem.tex
\section{Statement of the Problem} \label{sec:problem}

\Lead{Michael Blanton}
\Contributors{\ldots}



The current data ecosystem is the result of a large,  heterogeneous combination of efforts led by individual projects and missions, focused on achieving their specific scientific goals. With sufficient effort and resources, users have demonstrated that it is possible to combine disparate data sets and re-use some of the existing solutions.  
However, there are still several gaps, duplications, inconsistencies, barriers to access, productivity drags, missed opportunities for new capabilities, and a persistent danger of losing essential data sets. In this section, we summarize some of these problems. Currently, no existing organizations have the responsibility for resolving them.

Through the community, remarkable efforts exist to
mitigate the issues described in the section.
Examples are {\tt AstroPy}, the \gls{IVOA}
, the {\tt PyVO} client interface, and others.  Yet these sorts of
efforts overall remain underfunded and have not been able to resolve the
problems described here.

\subsection{The scientific user perspective}

Although the ability of science users to find, access, and use astronomical data has significantly improved over the past two decades, it has not kept pace with the greater computational, data storage, and \gls{software} infrastructure available today.

From the user's perspective, various archives distribute a rich assortment of astronomical data. However, finding the right archive or instrument requires perusing numerous websites, creating a barrier to access and leading to missed opportunities. Each archive has different authentication and authorization protocols, \gls{API}s, documentation systems, and file formats. Some of this heterogeneity is inevitable but, in most cases, the fundamental data sets are similar. The development of standards and tools for accessing the data has improved the situation, but combining even small data sets from multiple archives comes with a significant overhead  in time and effort.

Users face even more fundamental issues when their science requires large-scale manipulation of data across archives. Examples include executing a cross-match between two data sets with billions of rows, or performing an image calculation that requires touching a substantial fraction of pixels in an imaging data set. While some archives have server-side capabilities (e.g., science platforms) that allow large-scale manipulations within their own holdings, the lack of communication between the facilities is a serious obstacle to joint manipulation of data sets hosted in different archives or to sharing results with other users without downloading large amount of data.  These shortcomings represent a barrier to access, lead to both duplications and inconsistencies (since users develop similar but slightly different solutions to the same problem), and missed opportunities for new capabilities.

Similar issues also affect the availability of advanced software tools to analyze the data. Several specialized packages designed for common astronomical data formats and analysis tasks are available, but their maturity and support still leaves many gaps. As an example, over the past ten years, the Python {\tt AstroPy} package has become arguably the reference astronomy package. In that respect, it has replaced the older \gls{IRAF} package, primarily because the Python platform provides a much more common and general framework than \gls{IRAF}'s \gls{CL}. However, despite the tireless and often unrecognized effort of generous contributors, {\tt AstroPy} still does not have many of the image and spectrum analysis packages that the community needs.

\subsection{The project, observatory, and mission perspective}

Like the scientific users, organizations that generate astronomical datasets --- space missions, observatories, and projects that utilize them --- have not been able to take full advantage of the increases in computation, data, and software capabilities. Among other fundamental issues, that the structure of funding discourages an organized approach to coordinating data and software efforts across missions and archives. Individual projects are funded for specific tasks and products, and there is no incentive to develop shared solutions, except for rare cases where projects are funded by the same organization.

In some cases, software efforts associated with large project are subject to funding limitations, especially when the projects are faced with difficult budgetary decisions. This situation can lead to a vicious cycle, where software that is poorly documented or engineered becomes more difficult to reuse, to the detriment of future projects.

This stovepiped and underfunded situation has several deleterious effects:  new projects  cannot benefit from the knowledge, expertise, and tools generated by previous projects.  Projects also have little incentive or spare capacity to work together on the development of standards, or even to participate in a common forum to communicate about shared methods, problems, and solutions.

When it comes to archiving data, federally-funded projects have nominally ``natural'' federal archive homes depending on their funding source, but there is no established protocol to ensure that the archives have the resources to perform all of the archiving tasks. Quite often, new projects lead to new contracts or arrangements with the appropriate archive. While NASA offers a suite of organized solutions, this issue is particularly relevant for ground-based astronomy. 

For projects that are primarily privately funded (e.g., the Sloan Digital Sky Survey), there is no natural federal archive home, and the longevity of access to these datasets is not at all clear.

\subsection{The archive perspective}

The astronomical archives and the staff working at them face some of the same issues described in the previous two sections.

In some cases, archives have core missions that are defined in terms of specific data sets. When funding is associated with deliverables involving those specific data sets, archives lack flexibility to address issues related to combination of different data sets, sometimes even within their own holdings, and especially with other archives. Even participating in a common communication forum, jointly developing standards, and upstreaming generally useful software components might be hard to justify.

Incorporating standards can be particularly difficult because they can be complex and there are limited resources available to help archives with adoption. Additionally, standards may need modifications or extension to become useful for new projects, but modifying them thoughtfully requires longer timescales than project deadlines allow, and there is no long term set of requirements driving archive development funding. The lack of a natural way to fund cross-archive collaborations can lead to  unnecessarily redundant work. Opportunities to develop common tools and standards are not impossible to take advantage of, but they are harder to arrange and fund than necessary.

Modern technical advances, such as \gls{cloud} platforms, are difficult to explore because their cost is hard to predict. A greater culture of sharing information and planning would allow the astronomical archive community to explore these opportunities more effectively. However, this culture cannot grow without enough free energy among the archive scientists and without adequate incentives.

A related problem is that there is no clear career path for data or software-oriented astrophysicists within traditional academic settings or the data archives and national labs. This work is often defined as ``outside'' astronomy, even when it requires a deep understanding of the data sets and methods that all of astronomy depends upon. This situation makes it hard to define a career as a software-oriented astronomer/astrophysicist.

%% file: ecosys_short.tex
\section{The Astronomical Data and Software Ecosystem} \label{sec:ecosys}

\Lead{Adam Bolton}
\Contributors{Kim Gillies, Kerk Kee, Knut Olsen, Alberto Accomazzi, Bryan Miller,
Gregory Dubois-Felsmann, Curtis McCully (did we forget anyone?)}

\subsection{Data Ecosystem Introduction}


Software and data systems are both ubiquitous and foundational to astronomy in the 21st century \citep[][Section 4.5]{NAP26141}. The image of the lone astronomer squinting at the eyepiece during the night has been replaced by a modern system of digital detectors, observatory control \gls{software}, complex computerized data-processing algorithms,
and multi-petabyte-scale archives with rich online user interfaces. Major
trends such as the rapid growth of time-domain astronomy and the expanding
capabilities of artificial intelligence and machine learning have accelerated
this transformation. Discoveries at all scales, from the Solar System to
the most distant galaxies in the Universe, are made today within an
ecosystem defined by \gls{software} and data. This chapter presents a
high-level overview of the activities, elements, and participants within this
ecosystem. While the focus of \cite{NAP26141} is on archives and pipelines, we believe that a holistic  development of the entire ecosystem is the most effective approach.

\subsection{The cycle of discovery}

\begin{figure}
\begin{centering}
\includegraphics[width=0.8\textwidth]{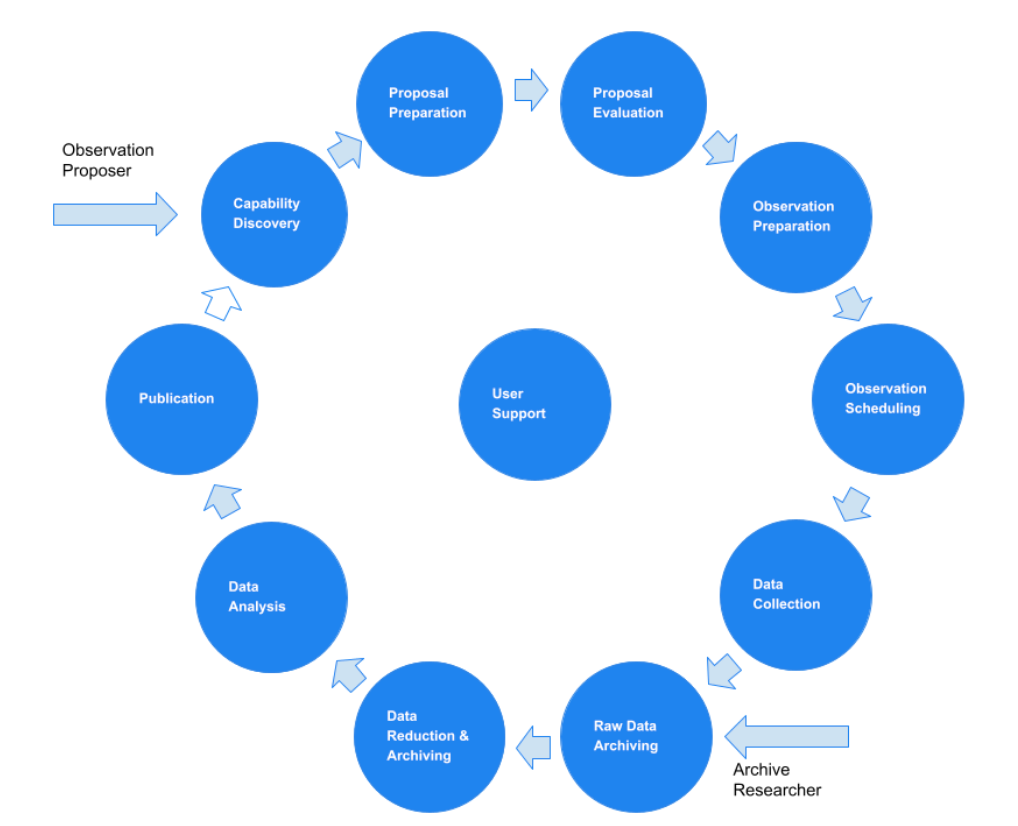 }
        \caption{
The ``cycle of discovery'' workflow for queue-based, \gls{PI}-led observational experiments. Proposals for new observations start with capability discovery. Other projects may start by scouring archives for existing data.
\label{fig:cycle}}
\end{centering}
\end{figure}

The experience of scientists within this ecosystem can be visualized  as a
``cycle of discovery'' illustrated  in Figure \ref{fig:cycle}, with three
archetypal variations or modes.

\begin{enumerate}

\item In the {\it \gls{PI}-driven mode}, scientists or teams of investigators propose research projects by applying for observing time on existing telescopes, which provide open access to the most meritorious proposals. For non-federal observatories or consortia, access might be restricted to scientists from select institutions or communities. Observatories vary in software and data support: in some cases, they offer platforms for proposal submission, automated pipelines for ready-to-use data, advanced analysis tools, and data archives, while some only provide basic infrastructure, requiring PIs to handle software and data management.

\item  In the {\it project mode}, large collaborations of scientists seek funding from federal agencies or private foundations to construct new telescopes, instruments, or space observatories for ambitious scientific projects that go beyond what's achievable in PI-driven mode. The team oversees the facility's construction, operation, and primary research. The necessary software and data infrastructure are developed within the project, sometimes with support from shared cyberinfrastructure like federally funded high-performance computing facilities. This approach aims to generate extensive and consistent scientific datasets, fostering opportunities for archival research.


\item In the {\it archival research mode}, scientists engage in research by mining previously collected data from archives, rather than seeking telescope time or funding for new experiments. The volume of archival data has grown, notably due to survey projects covering extensive sky areas, leading to a rise in discoveries from archival research. Since most astronomical archives are publicly accessible, this mode enables scientists from various affiliations to access research opportunities without institutional or collaborative constraints.
\end{enumerate}

Data archives and astronomical \gls{software} are the critical elements that support all of these three modes.

\subsection{Archives}

Astronomical data archives serve as the primary storage for astronomical data in the digital era. These archives are highly distributed across the current ecosystem, often associated closely with specific projects or observatories. Responsibility for different archives is usually divided by funding agency, with NASA overseeing space-based astronomy and NSF managing ground-based efforts in the \gls{US}. Multi-mission archives, particularly for space missions, are organized by wavelength (such as MAST, IRSA, \gls{HEASARC}). Private ground-based optical observatories offer archives with a range of services, from basic data preservation to comprehensive archives integrated with data pipelines producing advanced data products.


Archives face the challenge of balancing operational data management and \gls{PI} data access with support for archival research. These goals often demand different features, designs, and performance optimizations, all of which must be managed within limited funding.


The diversity and federation of current archives ensure responsiveness to the priorities of their respective communities. This also maintains a strong link between archival data and the observatory or project staff familiar with it. However, it presents challenges for research projects spanning multiple archives, especially those in different national communities. Standards for archive \gls{interoperability}, partially developed and implemented under the IVOA, help address this challenge by enabling data discovery and access through a common interface across multiple archives.

The advent of large survey projects with extensive and consistent datasets has created new challenges and opportunities not entirely covered by current models for archive interoperability. These datasets are too large for individual investigators to download, leading to the need for new approaches like data co-location and server-side data analysis in addition to search. This has prompted the development of astronomical "science platforms". For more details, refer to section on Science Platforms (\appref{sec:sciplat}).


\subsection{Software}

The elements of data infrastructure in astronomy are largely defined
and powered by \gls{software}. Different projects, observatories, and modes
of research have different \gls{software} requirements and architectures;
however, a number of common elements can be identified:

\begin{itemize}
    \item Proposal \gls{software} (“Phase I”): Observatories whose science program
    is driven by observing proposals submitted by individual investigators
    and teams typically employ a \gls{software} platform for proposal submission
    and processing. These systems include an interface for preparing and
    submitting proposals, as well as infrastructure for routing proposals
    to reviewers, collecting review results, communicating results to
    proposers, and forwarding successful proposals for further implementation
    and scheduling.
    \item  Observation planning \gls{software} (“Phase II”): Observatories that employ
    a queue or service-observing model to execute observations on behalf of
    investigators typically employ a \gls{software} platform that enables
    investigators to specify the details of their observations—e.g.,
    instrument modes, \gls{calibration} sequences, exposure times—in sufficient detail.
    See  \appref{sec:opsplat}. 
    \item Scheduling \gls{software}: Many observatories employ \gls{software} to build an
    observing schedule, either as an aid to human schedulers or as a more
    fully automated service. The nature of scheduling \gls{software} varies greatly
    according to the mode of research and the overarching science goals. For
    example, scheduling multiple \gls{PI}-driven programs poses a different problem
    than scheduling a single large-scale survey.
    See  \appref{sec:opsplat} 
    \item Observatory control \gls{software}: the physical operation of all modern astronomical
    observatories is computerized and controlled through \gls{software}. This \gls{software}
    controls the motion of large and expensive machines, and must ensure
    the safety of personnel and equipment.
    See  \appref{sec:opsplat}. 
    \item Instrument control \gls{software}: modern astronomical instrumentation is
    also controlled through \gls{software}, which is often developed and maintained
    separately from observatory controls. This distinction results from the
    common model under which instruments are delivered to observatories by
    independent or quasi-independent groups that maintain their own \gls{software}
    frameworks and standards.
    See  \appref{sec:opsplat}. 
    \item Data transfer and curation \gls{software}: all observatories operate some
    form of \gls{software} for management of their science data. At its most basic,
    this may only provide basic data-backup along with a mechanism for PIs
    to carry their data home from the telescope. At the other extreme,
    this may provide automated transfer to a full-featured archive or data
    center with multi-site backups, along with automated execution of
    data-reduction \gls{software} (below).
    \item Data reduction \gls{software}: all astronomical observations require \gls{software} to
    transform raw detector data into science-ready data products such as images, spectra,
    data-cubes, and catalogs. In some cases scientists configure and operate this
    \gls{software} themselves, and in other cases projects or observatories operate
    it in the form of pipelines. This class of \gls{software} encompasses both
    algorithmic methods and execution frameworks.
    See  \appref{sec:pipelines} 
    \item Data analysis \gls{software}: astronomers use \gls{software} to extract scientific
    measurements, results, and insight from reduced astronomical data. Software
    tools in this category (broadly defined) enable data visualization,
    exploration, model-fitting, physical parameter measurement, statistical
    characterization, and other forms of high-level analysis. This class of
    \gls{software} includes both data analysis pipelines for specific projects or
    instruments, and more general purpose astronomical \gls{software}.
    \item Data archive \gls{software}: in addition to providing fundamental data curation as
    described above, data archives provide \gls{software} interfaces (both interactive and
    programmatic) for data discovery and access to a project's or observatory’s data
    holdings. These interfaces are used both by primary investigators and by archival
    researchers. Data archive \gls{software} is also used by observatory staff to
    support ongoing stewardship of data and associated \gls{metadata}.
    See \appref{sec:crossarc} 
    \item Science platforms: As astronomical data sets have become too large to
    download, new online platforms have enabled astronomers to analyze the data where
    it lives. These ``science platforms'' typically provide Jupyter notebook
    interfaces, pre-installed analysis \gls{software}, methods to query and cross-match
    large astronomical catalogs, integration with data archives through
    flexible interfaces for data access (e.g., image-cutout services), and
    allocations of storage and computing resources to users. A more detailed
    discussion of science platforms is presented in Section \ref{sec:sciplat}
    of this report.
\end{itemize}

In the current ecosystem, most facilities develop and maintain their own instances of data infrastructure \gls{software}, though some libraries (e.g., CFITSIO, WCSLib) are often shared. The majority of current investment in astronomical software is channeled through specific projects and observatories, and the resulting development is directed towards the specific needs of a project.

Major exceptions to this trend have been concentrated in the areas of data reduction and analysis \gls{software}, where the development of reusable software tools has been more common. Historically, the US-based \gls{IRAF} project and the \gls{UK}-based Starlink project developed and supported
general-purpose \gls{software} for astronomical data reduction and analysis.  In the present day, many aspects of astronomical data analysis are supported through the open-source Astropy project and a large number of affiliated packages.

\subsection{Actors and Institutions}

The majority of astronomical data management and \gls{software} development in
the \gls{US} is funded by the three main agencies that fund astronomy and
astrophysics in general: \gls{NASA}, \gls{NSF}, and \gls{DOE}. NASA’s priorities are
driven by science requirements to support the missions and research
that it funds. \gls{NASA} has also recently begun to prioritize open science\footnote{https://science.nasa.gov/open-science-overview}
and open data access as agency goals. As a highly mission-driven agency,
software efforts funded by \gls{DOE} also are required to be well connected to its
major projects such as \gls{DESI}, Rubin, and \gls{CMB-S4}. \gls{NSF} is traditionally
a more grant-driven agency that also runs large astronomical facilities
with multi-decade lifespans, and it funds a correspondingly
heterogeneous range of \gls{software} and data activities.

While still a minority of total investment, private philanthropic
foundations provide increasingly significant support for astronomical
software and data systems. Particular examples are the \gls{LINCC} initiative
funded by the Schmidt Futures Foundation, various smaller projects
funded by the Heising-Simons Foundation, and the sponsorship of this workshop by the Simons Foundation’s Flatiron Institute. The
Sloan Digital Sky Survey (\gls{SDSS}), funded in part by the Alfred P.
Sloan Foundation and primarily funded by its member institutions, has
been developing and deploying innovative astronomical \gls{software}
and data systems for over two decades.

The development and maintenance of astronomical \gls{software} and data
systems is a collaborative effort between scientists and \gls{software}
engineers, including many people who fall on a continuum between
those two categories. Some are based at universities (faculty and
postdocs) and others are based at observatories, national labs, and
science institutes (staff scientists and engineering staff). Software
and data collaboration within and between institutions adds an
additional degree of complexity in the context of multiple large
projects spanning multiple institutions. Importantly, an individual’s
job description and career path is defined largely within the context
of their employment institution.

The computational infrastructure supporting astronomical \gls{software}
and data management is a diverse and evolving aspect of the ecosystem.
Traditionally observatories and their associated archives have operated
their own in-house computing systems to meet their needs. The role of
campus-level and national-scale research computing centers in
supporting astronomy data management has expanded over time, especially
for large \gls{DOE}- and NSF-funded survey projects. More recently, many
projects and centers have begun migrating their infrastructure to
commercial \gls{cloud} environments. Orchestration frameworks such as
Kubernetes have facilitated more flexible approaches to deployment
that are not tied to specific locations.

\subsection{Conclusion}

The ecosystem described above has evolved in a largely organic manner, without centralized coordination or decision-making. While this has led to great diversity, innovation, and responsiveness to immediate scientific needs, it has also resulted in significant duplication, incompatibilities, and difficulty in scaling to solve complex field-wide problems.
To address this, we need to build a common software ecosystem that meets the challenges of projects with specific requirements, individual schedules, big data challenges, processing time constraints, and scientific user communities with a range of use cases. This requires a change in the way we do business. To develop common software frameworks, we must analyze individual projects at a detailed level and across wavebands to identify areas in the software ecosystem that are common, and recognize areas where software needs to be specifically configured at the project level (but still plug and play).
The system must provide the flexibility that enables each project to select and customize their operational configuration at a high level. For data archives, metadata for data discovery and accessibility will need to follow IVOA standards, and scientific use cases will need to drive the definition of scientific metadata to ensure science analysis is supported downstream.
Most projects have been funded to focus on their own requirements and schedule. To achieve change at this scale, new ways of coordination and funding, as discussed in the following chapters, are required. Looking forward, the challenge for the astronomical software and data community is to preserve the positive benefits of the current ecosystem while addressing how to build a common software baseline.
There are clear opportunities to leverage modern technology, open-source and open-development software methodologies, greater coordination across agencies and institutions, and a diverse and capable professional workforce.

%% file: scope.tex
\section{A Roadmap: Scope, Requirements, Use Cases, \& Implementation Guidelines} \label{sec:scope}

\Lead{Janet Evans (janet@cfa.harvard.edu)}
\Contributors{Paul Barrett (pebarrett@gwu.edu), Tim Jenness (TJenness@lsst.org),
Jeff Kern (jkern@nrao.edu), Jorge Gonzalez Lopez (jgonzalez@carnegiescience.edu),
Dara Norman (dara.norman@noirlab.edu), Felix Stoehr (fstoehr@eso.org),
Kyle Westfall (westfall@ucolick.org)}

\subsection{Introduction to the Roadmap}

We present a roadmap using a system engineering approach to support
the recommendations included in the Astro2020 report \cite{NAP26141}.   Functional requirements frame the overall plan and apply as follows: Level 0 are the high-level project objectives that are mapped to guidance identified in the decadal report; Level 1 are agency level requirements and guide how the plan going forward is organized and implemented; and Level 2 requirements are flowed down to individual projects and are the basis of project standards and protocols.  We outline a several-phased approach to guide the agencies and community forward. The proposal includes a roadmap, a plan for implementation, and guidance for overall community support.

By engaging the community to define and socialize a set of
high-level engineering requirements (Level 1), a clear community
vision for achieving the goals defined in the Astro2020 report
can be established. It is important to note that these are not
technical solutions or approaches but rather statements of how
the system shall function. With these requirements as guiding
principles the efforts of individual projects will be aligned
toward the final state described in Astro2020.

Changes on the scale recommended in Astro2020 are challenging
and require time and effort to achieve but will have a positive
impact on current and future projects. The recommendations have
themes of efficiency in development and cost, trusted and
interoperable data products with well-defined \gls{metadata}, easy
data access for all the community (whether from large institutions
or small universities), and a call for common infrastructure and tools
in astronomy to enable research and science discoveries.  Doing so requires a concerted effort to develop these elements, to gain support from stakeholders in the community, in archive centers, and in agencies and foundations, and to build the relationships between those stakeholders.
As we meet future challenges with ever-more complex instrumentation and
petabyte and larger scale datasets, cost-effectively sharing
software and data facilities from technological and scientific
perspectives is a key challenge before us. We need to remind
ourselves not to break what is already working well in astronomy
and recognize that existing successful projects can help guide
the way into the future. It is our resolve that the biggest risk
is to do nothing at all and a more productive path is to make
strides to achieve the challenges of Astro2020. We offer the
start of a risk list below.

The start of a list of risks we see are:
\begin{itemize}
    \item Whole community is not included
    \item Lack of community vision of what we’re trying to achieve
    \item Unrealistic expectation management - what can be achieved in 1 year, 3
    years, 5 years, 10 years
    \item Level 1 requirements are insufficiently precise to drive verifiable
    Level 2 generation
    \item Feedback loop is insufficient for course corrections
\end{itemize}

\subsection{Scope}

The overall scope as guided by the Astro2020 report is extensive and the community (both federally
and privately funded) is asked to consider the entire flow of astronomical data from acquisition
at the telescope to access by the community.   Agencies and stakeholders are guided to address how to
develop data systems that consider the entire flow of data from the instrument to
pipelines producing science-ready data products, to archives
storing data in standard formats, to analysis, and publication,
and to make funding opportunities dependent on those capabilities.
The guidance includes direct language to \gls{NASA} and NSF to explore
mechanisms to improve coordination among U.S. archive centers and
to create a nexus for interacting with international archives,
further supporting the broad scientific needs of the astronomical
community at large.  Astro2020 also states ``Astronomy has entered an era in which well-designed and well-constructed software can be as important for the success of a project as hardware'' and that software developers and large software development are not adequately funded.   Petabyte datasets
are here, and the challenges of big data and \gls{software} management
will increase over the next decade.  The report goes on to guide
decreasing barriers to \gls{HPC} and \gls{HTC} computing, flexible career paths for scientists
and \gls{software} developers, and guidance to universities regarding
increased data science and machine learning curricula.

\subsection{Requirements}

Table \ref{table:l0l1reqs} includes \gls{L0} and example \gls{L1} requirements.  L0 are derived from Astro2020, and
L1 are a flow-down from the L0 requirements.  An important measure
to gauge the level by which the community embraces the approach
and to gain buy-in is to involve the astronomical community
in the definition and refinement of requirements.

\startlongtable
\begin{deluxetable}{rrr}
\tablecaption{L0 and \gls{L1} requirements examples} \label{table:l0l1reqs}
\tablehead{\colhead{Level} & \colhead{Requirement} & \colhead{Source}}
\startdata
L0 &
\begin{minipage}{\midcolwid}
\vspace{5pt}
Progress [in creating effective data archives] will come from an end-to-end approach that considers the entire flow of data from the instrument, to the archive, to analysis, and publication.
\vspace{5pt}
\end{minipage}
& Astro2020 \cr
L0 &
\begin{minipage}{\midcolwid}
\vspace{5pt}
The \gls{NSF} and stakeholders should develop a plan to address how to design, build, deploy, and sustain pipelines for producing science-ready data across all general-purpose ground-based observatories (both federally and privately funded), providing funding in exchange for ensuring that all pipelined observations are archived in a standard format for eventual public use.
\vspace{5pt}
\end{minipage}
& Astro2020 \cr
L0 (suggested) &
\begin{minipage}{\midcolwid}
\vspace{5pt}
NASA and \gls{NSF} should explore mechanisms to improve coordination among U.S. archive centers and to create a nexus for interacting with the international archives communities. The goals of this effort should be informed by the broad scientific needs of the astronomical community.
\vspace{5pt}
\end{minipage}
& Astro2020 \cr
L0 (suggested) &
\begin{minipage}{\midcolwid}
\vspace{5pt}
Software should be publicly available to minimize redundancy, incentivize the adoption of common standards, and promote applications using multiple data sets.
\vspace{5pt}
\end{minipage}
& Astro2020 \cr
L0 (suggested) &
\begin{minipage}{\midcolwid}
\vspace{5pt}
NASA, NSF, and the national facilities should strive to decrease the barrier to entry to those who want to engage in \gls{HPC} and \gls{HTC}.
\vspace{5pt}
\end{minipage}
& Astro2020 \cr
L0 (suggested) &
\begin{minipage}{\midcolwid}
\vspace{5pt}
Establish flexible and stable career paths for archive scientists and \gls{software} developers to preserve the expertise and resources of the existing data centers, and to support the development of scientific \gls{software} infrastructure and open-source \gls{software} projects.
\vspace{5pt}
\end{minipage}
& Astro2020 \cr
L0 (suggested) &
\begin{minipage}{\midcolwid}
\vspace{5pt}
Data science and machine learning should be incorporated into graduate level and beyond training to prepare researchers for its increasing role in astronomical research.
\vspace{5pt}
\end{minipage}
& Astro2020 \cr
L0 (suggested) &
\begin{minipage}{\midcolwid}
\vspace{5pt}
Universities should establish hybrid teaching and research tracks that combine research and \gls{software} development similar to those for research and instrument development.
\vspace{5pt}
\end{minipage}
& Workshop \cr
\hline
L1 (suggested) &
\begin{minipage}{\midcolwid}
\vspace{5pt}
An incentive and funding structure that favors collaboration and code-reuse over writing from scratch.
\vspace{5pt}
\end{minipage}
& Workshop \cr
L1 (suggested) &
\begin{minipage}{\midcolwid}
\vspace{5pt}
Well structured, well maintainable \gls{software} base to address the problem, designed from the start to be (or extended to be)  multi-mission.
\vspace{5pt}
\end{minipage}
& Workshop \cr
L1 (suggested) &
\begin{minipage}{\midcolwid}
\vspace{5pt}
A set of multi-mission, documented interfaces and implementations supporting each of the stages in the circle of discovery; These may evolve and support existing systems and current missions.
\vspace{5pt}
\end{minipage}
& Workshop \cr
L1 (suggested) &
\begin{minipage}{\midcolwid}
\vspace{5pt}
An established methodology to promote the prototyping, development, evolution, and adoption of implementation of processing stages.
\vspace{5pt}
\end{minipage}
& Workshop \cr
L1 (suggested) &
\begin{minipage}{\midcolwid}
\vspace{5pt}
The system shall coordinate the storage of data in long-term archives, and identify cases where long-term storage solutions do not exist yet.
\vspace{5pt}
\end{minipage}
& Workshop \cr
L1 (suggested) &
\begin{minipage}{\midcolwid}
\vspace{5pt}
The system shall support the transport and maintenance of metadata throughout the cycle of discovery.
\vspace{5pt}
\end{minipage}
& Workshop \cr
L1 (suggested) &
\begin{minipage}{\midcolwid}
\vspace{5pt}
The system shall support interoperable data.   [one such tactic could be a standard data model].
\vspace{5pt}
\end{minipage}
& Workshop \cr
L1 (suggested) &
\begin{minipage}{\midcolwid}
\vspace{5pt}
Define the interface between the nodes in such a way that alternative solutions can emerge without destroying the end-to-end \gls{stack}.
\vspace{5pt}
\end{minipage}
& Workshop \cr
\enddata
\end{deluxetable}

\subsection{Use cases }

Use cases are an important driver of any science process.  Here, too,
a challenge is to gather use cases from the community
that span the range of expectations of the program, data archives, and
data systems to understand the ecosystem clearly.  There is room to add
use cases over time, but a better initial effort ensures that the design
is more stable.

Use case examples:

\begin{itemize}
   \item {\bf Example 1:} In 2030, a proposal comes in to build a new general purpose
   ground based telescope. How can this plan facilitate the goal of having a large fraction of its end-to-end \gls{software} drawn from (or built upon) existing code bases? Progress towards this goal can be measured against existing missions, or against potential future ones (\gls{US}-ELTP, MegaMapper, etc.).
   \item {\bf Example 2:} An independently developed \gls{software} package
   wants to upstream its functionality to the wider community.  How does this plan  facilitate this?
   \item {\bf Example 3:} A new instrument in early stages of development is
   starting to think through their data-management systems. How does this
   plan help link the instrument team to existing resources; e.g., instrument
   control systems, data reduction and analysis pipelines, data archiving
   solutions/facilities, science platforms for users?
   \item {\bf Example 4:} A small observatory has a mandate from its users/funders
   to deploy a community-facing archive. How does this plan enable smaller
   observatories, with limited funding/personnel, to build that archive
   successfully, including the ability to host proprietary data with
   authentication protocols?
\end{itemize}

\subsection{Implementation}

The roadmap we present can be implemented in a multi-faceted and staged
approach.  Key components for agency support and guidance include archives
and data systems.

We acknowledge and accept the reality that along with a multi-faceted and staged
approach, course corrections to any plan are expected moving forward over
time.  Targeted evaluations that consider changing community needs and
responses may identify needed adjustments to ensure that the plan produces
software and products that are in line with the recommendations of Astro2020
over the decade.  Examples are measuring project success against the proposal
goals and evaluating whether projects are within budget (cost and time).
Occasional surveys throughout and at the end of each project will help to
collate and evaluate the project team's experiences.  A measure of synergy
may consider how well \gls{software} produced by a group or groups is subsequently
taken-up and re-used by the community or another project.

We have defined several phases to implement a plan that can be summarized
as follows:
\begin{itemize}
\item {\bf Phase 0:} Define a Roadmap (this chapter---framework, discussion, examples) and achieve buy-in from funding agencies.
\item {\bf Phase 1:} The NSF and \gls{NASA} empower and fund a small but representative steering committee through an open nomination process to:
\begin{itemize}
    \item Lead development of Level 1 requirements;
    \item Define a governance model for the maintenance and implementation.
    \item Develop use cases with the community to drive the processes.
\end{itemize}
\item {\bf Phase 2:} Agencies enact the implementation plan delivered in Phase 1.  Measure and adjust the plan.
\end{itemize}

Phase 0 defines a roadmap (this chapter) and charges specific tasks to
subsequent phases of the process.  In order to succinctly communicate to the funding agencies,  a light weight Letter of Support that has widespread buy-in among important signatories can be circulated to agencies as a justification for the actions expressed in this paper. The Letter of Support draws for justification on the charge in Astro2020, on Petabytes to Science, and on this workshop with experience, models, and this Roadmap.  A framing event for releasing the Letter of Support is  the designation of 2023 as A Year of Open Science.
The final paper and the Letter of Support will be circulated among and advertised to a broad set of stakeholders. The key parties are government funding agencies, private foundations, observatory directors and boards, the Astronomy and Astrophysics Advisory Committee, and other influential actors in U.S. astronomy. The goal is to gain the financial, administrative, and moral support of these organizations. Outreach will be in the form of individual contacts but may include gathering representatives of these organizations together.

Phase 1 empowers a small, representative  steering committee with support from community members and the funding agencies to develop the Use Cases and Level 1 requirements based on community input and broad acceptance as part of  the process.  The Level 1 requirements are imposed on future projects implementing this plan unless explicitly given relief by the agencies.  A second responsibility of the steering committee is to develop an implementation plan to achieve the
Level 1 requirements.  The implementation plan includes defining a governance model that encourages community participation as well as plans for long-term administration of the model.  A new project implementing this plan proposes to implement components that are part of the astronomical ecosystem and deliver Level 1 and Level 2 requirements that are defined as part of the proposal. A primary, underlying principle moving forward  is that we do not undermine or delay progress on current archives and activities or other arising opportunities.

Detailed guidance in fulfilling Level 1 and Level 2 requirements will need
to consider many aspects.  For example, can implementation of Level 1
requirements be phased in over several iterations of a component by the
same or different projects/proposals?  How will Level 1 and Level 2
requirements be verified?  How will \gls{software} be verified to meet contractual
obligations?  Are there design and code reviews, test plans and
verification, review/feedback/support/contribution (i.e., involvement)
from the original component development group?  Will there be a framework that focuses on data formats, standard protocols, standard API endpoints, documentation and development guidelines  or
will development be open and guided to follow that of the original group?
Does a component get re-used organically or is there a hurdle to jump
through to get it on a ``candidate for re-use'' list of some kind?
Will regression tests from the original and subsequent projects follow
a component to assure there are no regressions in commonly used \gls{software}?
Is it a requirement to have those tests pass?  Will there be guidance
on portability platforms to assure the most popular platforms in
astronomy are covered?  Does \gls{software} maintenance (e.g., Python,
cfitsio version) occur based on proposal rounds or are projects given
funding to support the infrastructure and maintenance up front?

Another consideration for the steering committee is to determine the models
to stand up to manage astronomy \gls{software} and data centers.  For example,
one could imagine a management model for archival data centers whereby
projects contribute data products following \gls{interoperability} standards
to an assigned center (or centers).  A different model may be more
appropriate for data systems (e.g., proposal planning, pipelines) that use common infrastructure components or implement large applications that are more appropriately managed using
a distributed approach allowing for flexibility but providing well defined
interfaces/standards.  For any implementation plan, carefully tracking
progress with appropriate metrics will guide any course corrections needed
to keep on a positive path.

We strongly recommend that steering committee members are drawn from a wide
cross-section of the communities/projects in astronomy (e.g., large,
small, new, and old) through an open nomination process.  To achieve acceptance, our guidance is to share, present, and receive feedback from projects (stakeholders) through various forums.  As examples, the steering committee should hold community listening sessions, include agencies and targeted informed members
of the community for review, and present at sessions of the \gls{AAS} meetings
to engage and receive feedback from the wider community.  We envision
an iterative process in Phase 1 where feedback is folded in and then
presented back to the community in accessible documents and/or in
subsequent sessions as needed.

Several models are outlined Chapter \ref{sec:casestudies} that
were developed as part of the workshop.  The group was asked to consider
if a model that called for a centralized agency was the way to go, or
if a federated model would be more realizable and successful, or if
a novel approach would make the most sense.  The attendees also
considered whether it is a combination of two or more of those models
that would be most feasible.
Four models emerged (Small Coordinating Body, Inter-agency Program Office, Common Engineering Pool, The Centralized Virtual Institute)
with an intent to inform the agencies and steering committee.  The next step
(and part of this roadmap) is for the Steering Committee to consider the pros and
cons of each, decide what is realizable or not, decide if there is a
singular or multi-faceted approach, and then choose one or more models,
or combine models into new/different models to go forward.  Perhaps
too, the Steering Committee will start from scratch and just use the models
described here to inform their discussion.

\subsection{Conclusion}

Early on, our vision is that individual development projects that are developing common solutions may not satisfy all of the Level 1 requirements, but they will satisfy a
defined subset of Level 1 requirements consistent with their projects’
needs and schedules.  The first subsequent project to utilize a common component
for reuse may have the most challenging task of all.  Developmental
prototypes may be the first step in evaluating whether a common component meets
a project’s needs and are a path towards evaluating whether the original
component was built in such a way as to be extensible.  Full compliance
with all Level 1 requirements may only be realized after several
implementations of a common component are completed by the same or different
projects.  With this vision in mind and over the next decade, common components
built to the full set of Level 1 requirements will emerge that enable
projects to share a common \gls{software} base.  Some components may be
project-specific or may need project-specific enhancements interfaced
to common infrastructure.  Even in these cases, modular, extensible,
and maintainable \gls{software} should use common standards and programming
practices, adhering to many of the Level 1 requirements.

As we look forward, the roadmap and considerations we outline in this
chapter are meant to take the community and funding agencies a long way
toward achieving the goals of Astro2020.  In addition, and as part of
the roadmap we outline, we believe community acceptance is a requirement
for success along with equity and inclusion for all participants and
projects (big or small).  We expect course corrections will be needed
as the phases are implemented and experience is gained over time.
The reward, if implemented correctly, is that astronomy computing will
benefit from an extensive data system vetted, curated, and  shared across
projects, groups, and individuals.   Moreover, the system will be
more technologically advanced, interoperable, and maintainable than
any one group could reasonably achieve.  The agencies will meet
the goals of Astro2020 at an overall cost savings, and the scientific
community will gain from an extensive and stable common \gls{software} base
across astronomy to shepherd research.

%% file: casestudies.tex
\section{Case Studies on New Structural Models} \label{sec:casestudies}

\Contributors{Adam Bolton, Vandana Desai, Gregory Dubois-Felsmann, Curtis McCully, Dara Norman, William O’Mullane, Adrian Price-Whelan, Luca Rizzi, John Swinbank}

\subsection{Introduction to Case Studies}


The astronomy and astrophysics ecosystem is broad. There are many different organizations and individuals with responsibilities for providing software that include all phases of the cycle of scientific discovery, from planning observations, to executing those observations on both ground and space-based telescopes, to reducing and analyzing the collected data and making higher level data products readily available, and so on.
Section \ref{sec:ecosys} (the data and software ecosystem) describes these groups and the various types of software in more detail.
The Astro2020 Decadal Survey challenges all participants in this ecosystem, and especially those controlling the flow of resources, to align their efforts to better support the community of researchers.

This section contains a few case studies as potential new structural models that address the challenges faced by the current astronomical data and software ecosystem.
These models are meant to serve as ``straw man'' plans to strategically evolve the current ecosystem to more effectively support the community and therefore achieve what was envisioned in the Decadal recommendations.
Each model has strengths and weaknesses, addressing some areas of concern well, while not effectively addressing others, or even unintentionally causing new areas of concern.
To address the full scope of concerns, it may be necessary to draw from more than one of the models discussed below, or to try more than one of these models simultaneously.

Models 1, 2, and 3 represent a sequence of increasing scope and responsibility for the structures considered.
Model 1 is a purely coordinating and advisory body with no means to disburse funding itself and no direct responsibility to actively work on the software or data ecosystem solutions.
Model 2 would establish a new program office to identify and fund cross-agency or other efforts that support the broader astronomical data and software ecosystem.
Model 3 defines a new physical institute to provide a stable workforce to implement and support the ecosystem.
As an alternative to these three, which might all have to exist to fully meet the needs of the community, Model 4 proposes an alternative solution to the combined Model 1-2-3.
Model 4 would create a new overseeing body (similar to Model 1) who would define the end-to-end data ecosystem and solicit bids from stakeholders to contribute sub-elements of the system (similar to Model 2).

In making any structural changes to address the concerns raised in the Decadal report, we advocate that initial implementations
should be set up as experiments with finite and realistic durations, resource allocations, goals, and measures of success.
This is a departure from how solutions to structural problems have been implemented previously, where a singular model is built as the solution, fully implemented, and then tweaked as problems arise.
We instead advocate here for a more agile approach to bringing about the conditions needed to more fully address the Decadal Survey recommendations around aligning software efforts across the astronomy ecosystem.
Starting with experimental or fixed-term implementations of any new structural models allows for established elements of the ecosystem to determine how to work with or within the new model; This will be crucial for community buy-in and to maintain continuity for users and other stakeholders as we move towards a more cooperative final model.

We anticipate that the costs to support these models might range from a few to a few tens of millions. In Appendix \ref{sec:costs}, we have attempted to provide a rough  cost estimate for each model, for comparison of scale across models more than as actual costs. Most costs are personnel based, with a single fully burdened estimate of \$250k used for all FTEs.
This may be a low estimate since personnel for many of these models will probably be heavily weighted to senior staff.
The number of personnel needed for each model also has a wide range.  We have used the FTE count scale to provide a low and high cost estimate multiplying by the FTE cost.
Hence, each model has a wide cost range reflecting the preliminary approach used in our costing estimate.  We acknowledge that selection of any model should not be made without a detailed plan and full costing estimate.

%% file: model1.tex
\subsection{Structural Case Study: Model 1 --- Small Coordinating Committee} \label{sec:model1}

\textbf{Summary:} This model proposes establishing an inter-agency coordinating committee. This committee would offer guidance to funding agencies and stakeholders in awarding grants for critical infrastructure development. Specialized working groups with in-depth expertise would be formed to assist in prioritizing and allocating funds. Committee members should represent diverse perspectives, and membership rotation would occur on a multi-year (3--5 year) basis.


The primary objective of the committee would be to identify and prioritize crucial \gls{software} infrastructure development for the broader astronomical community's benefit,  and facilitate and empower collaborations among teams funded by various independent agencies or projects.



The committee is distinct from existing advisory committees (e.g., the Astronomy and Astrophysics Advisory Committee; \gls{AAAC}) in that it would focus on the software and data ecosystem and liaise directly with stakeholders and the broader community.
It could exist as a subcommittee of the \gls{AAAC}, or a parallel committee that interfaces with the \gls{AAAC}.

This committee would:
\begin{enumerate}
\item Interface with the community by convening town halls, soliciting white papers, or through other open channels,
\item Advocate to the agencies and other funding sources for the ideas proposed by the community,
\item Create and propose new incentive models to support collaboration across projects and institutes, and advocate for buy-in from relevant stakeholders,
\item Facilitate opportunities for collaboration by organizing meetings and workshops aimed around the \gls{software} ecosystem and data infrastructure,
\item Identify priorities as part of the development of the \gls{software} ecosystem,
\item Support individual projects in navigating the funding landscape, especially in support of shared or community-oriented projects.
\end{enumerate}


\subsubsection{Opportunities with this Model}
\begin{enumerate}
\item It builds on the desire of many existing stakeholders to work together by creating a committee that works across projects and institutions.
\item It takes advantage of the expertise already present at individual institutions.
\item It works within existing funding structures, by adding a layer of coordination. It creates a new committee but not a completely new organization.
\item It can be effective in dealing with agencies, as it closely resembles existing models such as the \gls{AAAC} and other advisory bodies.
\end{enumerate}

\subsubsection{Risks of this Model}
\begin{enumerate}
\item Without the endorsement and commitment of the community and leadership of stakeholder organizations (e.g., agencies, observatories, and data centers), achieving the objectives of this model would be unattainable. The effectiveness and authority of this coordinating body hinge on the collaboration of current stakeholders. An initial strategy to address this could involve aligning stakeholder performance in future projects with the guidance and criteria established by the coordinating body.
\item Careful selection of membership is crucial to ensure the community's needs are adequately represented. One mitigation strategy could involve adopting a community-driven approach for member selection, coupled with a well-defined rotation schedule.
\item Introducing an additional layer of reporting and bureaucracy is a potential concern. To mitigate this, individual project teams could be insulated from the requirement to report directly to the coordinating group, with this responsibility falling under the purview of center management instead.  
\item Lack of a clear incentive for participation is a valid concern; individual stakeholders may prefer to work independently. A mitigation strategy could involve establishing a tangible reward system for involvement in the coordinating body. For instance, offering unrestricted access to products developed through collective efforts could serve as a compelling incentive.
\end{enumerate}


\subsubsection{Example scenarios enabled by this model}

\begin{enumerate}
\item A developer at Institution A seeks to utilize a complex \gls{software} tool from Institution B. The most effective approach would be for A to engage a developer from B, who is already proficient with the tool. However, funding barriers and administrative overhead make this process overly complex. The coordinating body could streamline this exchange by assisting both organizations in navigating the funding environment, drafting Memorandums of Understanding (MOUs), and offering incentives to both A and B. This support would enable A and B to demonstrate to their respective funding agencies the significance of the collaborative product to the community.
\item Members of the coordinating body could be invited to sit in mission and project reviews, to ensure that the right balance is achieved between using/building upon existing solutions and the need to develop new tools. They could also help to quantify the risk of introducing dependencies when re-using existing solutions.
\item Nobody modernized the \gls{NASA}-\gls{GCN}, meaning that subsequent surveys ended up creating new solutions. The coordinating body could have facilitated the modernization of the GCN by providing an avenue where the issue was raised, by facilitating the discussion among the interested parties, and providing a way for the stakeholders to communicate to their funding agencies.
\end{enumerate}

\subsubsection{Similar Existing Structures}

The \gls{NAVO} represents an existing coordinating body that has demonstrated aspects of this model. As an example of the types of activity that the \gls{NAVO} enabled: \gls{HEASARC}, \gls{IRSA}, and \gls{MAST} wrote a proposal with a budget to implement a limited and uniform set of \gls{IVOA} protocols. \gls{NASA} took some money out of the archive operating budgets to fund the NAVO proposal. The leadership of the archives coordinate with each other and empower the developers to work together on software, such as \texttt{PyVO} and \texttt{Astroquery}. The \gls{HEASARC} lead is the official project manager, but does not control the budgets or supervise developers at the other archives.

%% file: model2.tex
\subsection{Structural Case Study: Model 2 --- Inter-agency Program Office} \label{sec:model2}

\textbf{Summary:}
This model entails establishing an inter-agency program office or an independent endowed foundation dedicated to supervising policies and initiatives aimed at fostering a thriving astrophysics data and \gls{software} ecosystem. The program office would allocate resources and oversee endeavors focused on shaping policies, frameworks, and incentive structures that maximize scientific output for a wide audience within the context of the data and software ecosystem. This office could operate independently of existing organizations, with its funding source being immaterial to its disbursement capabilities. For instance, it could initially rely on private funding. Additionally, the model would necessitate the formation of an advisory or steering committee comprising community leaders.

This model, while similar in goal to Model 1, encompasses a broader scope and higher cost. In this case, the office not only provides support for critical software infrastructure development but also has the resources to directly fund these efforts, rather than solely offering advisory services.

The rationale for adopting this model stems from the recognition that advancing the data and \gls{software} paradigm in astronomy requires a diverse range of solutions. These include coordinating existing endeavors, allocating small engineering grants, establishing new prize fellowships or faculty support channels, and creating career paths. Given that not all of these approaches can be fully understood or predicted from the outset, and not all will achieve success, a top-down approach carries inherent risks.

Instead, an office equipped to fund a variety of programs, with community input and feedback, offers a means to continually identify and refine practices and solutions that prove effective. This approach allows for the organic emergence of best practices from the ground up, driven by innovative ideas and projects within the community.

The program office would:
\begin{enumerate}
\item Have paid staff to organize effort, think about programs, run funding calls, and facilitate collaboration, including both technical and administrative staff to help bring projects together,
\item Develop, issue, evaluate, and administer funding programs on several scales to enhance the state of \gls{software} and data in astronomy,
\item Facilitate and incentivize collaboration among externally- or internally-funded groups (e.g., funding to two or more groups could be contingent upon collaboration),
\item Explicitly fund and facilitate maintenance of projects and contributions that generalize existing functionality to benefit a broader community, and
\item Establish and track metrics to understand the impact of funding and to define ``success'' of the grant programs.
\end{enumerate}

\subsubsection{Opportunities with this Model}
\begin{enumerate}
\item A multi-agency staffed office has the potential to dismantle barriers.
\item This entity could enhance community activities that drive higher scientific output.
\item The primary objective is to create and implement dynamic initiatives/programs fostering \gls{software} and infrastructure within the community, emphasizing streamlined reporting due to the experimental nature. The focus is on continual evolution based on effectiveness, with a key aim being to discern successful strategies.
\item Across all models, there is potential to transcend agency boundaries through unified coordination of astronomical efforts by a cohesive body.
\item Could fund many small programs and continue to fund successful ones.
\end{enumerate}

\subsubsection{Risks of this Model}
\begin{enumerate}
\item Inadequate upkeep or a lack of long-term vision for the program office could hinder its ability to effectively recognize and reward demonstrated progress and success.
\item Without being sufficiently up to date or lacking a long-term vision, the program office might fail to influence projects early enough in their life cycle.
\item Failing to receive current demographic information on usage could lead the program office to overlook the need to discontinue programs/projects that are not proving successful. This could result in funds being allocated to projects with minimal users, potentially harming the office's reputation.
\item Without strong support and promotion from various agencies through inter-agency cooperation, the program office may struggle to achieve its goals due to a lack of authority.
\item If the community does not feel a sense of ownership over the programs and processes originating from the program office, it may perceive the office as just another reporting agency.
\item If the program office is (or is perceived to be) too biased and is believed to favor certain grantees or approaches without adequate justification, it will lose credibility within the community.
\end{enumerate}

\subsubsection{Example scenarios enabled by this model}

\begin{enumerate}
\item Set up an agile funding cycle for small \gls{software} projects  - potentially quarterly calls. (\gls{XRAC})
\item Fund several \glspl{FTE} on relatively long timescale potentially at several institutes for support of community \gls{software} and allow them free energy for experimentation (in the group of institutions).
\item Initial funding for faculty lines to bootstrap astronomical \gls{software} research as a discipline
\item Enable and support program to support coordination and participation within international organizations (e.g. \gls{IVOA} travel, etc.).
\item Enable and support sabbaticals between industry, archive centers, national labs, and related organizations.
\item Provide support to incentivize coordination for the projects, through funding or \glspl{FTE}.
\item Enable and support other tasks like long time data storage that individual centers are not incentivized to do that this entity could get this done (large project).
\item Fund an institute or project that can take ownership of the common aspects of the work, with people specifically hired to work for this entity, who would be sent to work on specific projects for a certain time. (See Model 3)
\item Provide bridge funding for data management and \gls{software} experts during periods of transition between projects to retain expertise.
\end{enumerate}

\subsubsection{Relevant Existing Structures}


Program offices at existing funding agencies (e.g., \gls{NSF}, \gls{NASA}, etc.) --- especially those that work between domains or divisions --- demonstrate the potential of this model.

%% file: model3.tex
\subsection{Structural Case Study: Model 3 --- Common Engineering Pool} \label{sec:model3}

\textbf{Summary:}
This model envisions creating a new organization to support the ``core'' astrophysics software \gls{stack}.
The organization would foster collaboration between institutions, projects, and researchers throughout the astrophysics ecosystem and provide a software engineering workforce to help develop and maintain community tools.
This would consist of research software engineers, project managers, and support staff who would be employed by the organization.
One component of this organization would be a centralized software engineering and project management team at a new or existing institution. This team would be available for consulting or contract work.
A second component would be a network of software engineers embedded within other projects and institutions.
Together, this organization would provide a means to identify cross-cutting functionality, to support code re-use and collaboration, and to contribute generally-useful tools and algorithms from project-specific applications ``upstream'' to community-oriented tools.
We envision a minimum of 30 software engineers in the central team, and at least another 15 software engineers distributed among projects, missions, universities, and institutions.
The organization should be fully funded and not required to apply for grants or other financial support.

The main goal of this model is to create an organization that can work between and outside the lines of existing projects and institutions to create the community infrastructure we need for a flourishing data ecosystem.
This includes advocating for code re-use, implementing shared standards, and providing support for following best practices for software development and maintenance.
The organization would have a mandate from funding agencies that they endorse the example standards and best practices used by this group, i.e. by aligning funding for projects by working with this group and/or contributing resources to it.
This organization may not have any grants of its own to give out but it can contribute support to projects through people and contracts.
In this way, the organization could be entirely proposal-driven, or could  independently decide who it architects and maintains its core services, or it could be something in between.

This model directly develops engineering and architectural consistency across the community by developing and applying expertise.
Each new project that draws on resources and expertise from this common pool is not starting from scratch.
In this way, the model is designed to develop similar and compatible infrastructure in different projects and to prevent unnecessary duplication.

The ``central + distributed'' nature of this organization also provides an opportunity to create new sustainable career paths for research software and data engineers in astrophysics.



The engineering group would:
\begin{enumerate}
    \item Define, develop, and maintain the core \gls{software} infrastructure upon which the entire ecosystem can build, reducing duplication and upholding engineering standards.
    \item Respond to community requests to extend that ecosystem to meet emergent needs.
    \item Provide architectural advice and foster software engineering best practices across the community.
    \item Develop standards for the components of an architecture to be used across projects, in close collaboration with those projects.
    \item Provide consulting to funding agencies by reviewing proposals for consistency with the wider ecosystem and lack of duplication.
    \item Participate in project reviews with funding and oversight agencies.
    \item Provide new, sustainable career paths for software engineers.
\end{enumerate}




\subsubsection{Opportunities with this model}
\begin{enumerate}
\item Provide training on applicable \gls{software} development and architecture.
\item Organize focused workshops on specific aspects of the astronomy \gls{software} stack.
\item Run a proposal process by which effort is allocated to specific core infrastructure components.
\item Provide \gls{software} engineer \glspl{FTE} as a  reward for agreeing to use standards and standard \gls{software}.
\item Demonstrate a transparent process around system architecture; promote community buy-in by providing support and good starting points not just an agency mandate.
\item Develop a proposal handling system for use in \gls{US} ground based missions (stretch goal also space missions).
\item By giving engineering support smaller centers and institutions can have technologically advance solutions which may otherwise be considered too complex.
\item Support existing tools such as Astropy and integrate them in solutions/standards implementations.
\item Could become maintainer of some community \gls{software}.
\end{enumerate}

\subsubsection{Risks of this model}
\begin{enumerate}
\item If institutions (e.g., NASA Goddard, the eScience Center at the University of Washington) that already have such groups fail to cooperate and instead choose to compete, then this could  undermine the model.
\item If no shared ownership emerges, then there will be no buy in from the community or existing institutions.
\item If agencies will not support this ``institute" model sufficiently, it could be  undermined by support for competing common efforts.
\item If \gls{Archive} centers are not properly integrated and/or are put in a competitive
relationship with this organization, the organization will not achieve its goals.
\item If there is a perceived lack of control of the engineers in the organization by the community, calls for the funding to go directly to the institutions could  undermine the enterprise.
\item If it  takes on the responsibility of maintaining \gls{software} produced by the projects, then it could become overwhelmed with outdated \gls{software} used by one or very few projects.
\item If coordination and communication are not strong in the organization, misaligned priorities with between partners could result in delays and personnel problems.
\end{enumerate}

\subsubsection{Example Efforts Enabled by this Model}
\begin{enumerate}
\item A PI at small institution could make use of the software pool on  medium-scale projects
\item Community software projects could leverage and donate workforce effort for coordinated projects
\item Partnership institutions could draw on expertise that they don't have in-house to complete projects and in return donate FTEs in other areas.
\end{enumerate}

\subsubsection{Relevant Existing Structures}

The UK Starlink Project was a UK astronomical computing project that supplied general-purpose data reduction and other software to astronomers until it was ended in 2005.
Starlink helped to standardize the way that astronomical data was reduced and analyzed while it operated and helped to train a new generation of astronomers in the use of astronomical computing.
Starlink demonstrated the utility of having an organization in charge of maintaining core infrastructure for the astronomical data ecosystem.

Another relevant institute is the (contemporary) Netherlands eScience Center.
The Netherlands eScience Center is a research organization in the Netherlands that collaborates with astronomers and other scientific disciplines to advance research through digital technologies.
The eScience Center develops research software for groups who apply for developer time.
By providing tailored support and training to researchers across many domains, the eScience center provides much-needed software development support to research teams and establishes new career paths for software-oriented scientific researchers.

The Rubin In-kind program includes a developer pool that can be deployed to projects within the Rubin ecosystem.  Although this is not a ``stand alone'' facility, some aspects of the pool's operation might be useful examples for this model.

%% file: model4.tex
\subsection{Structural Case Study: Model  4 --- The Centralized Virtual Institute} \label{sec:model4}


\textbf{Summary:} As mentioned previously, Models 1, 2, and 3 provide a sequence of structural models of increasing scope that aim to support the software and data ecosystem either through coordination, new funding, or a new workforce.
In practice, all three or some combination of these three models may be necessary to fully address the issues raised in the Decadal report.
This model --- the centralized virtual institute --- presents an alternative to the combined Models 1-2-3.

The new virtual institute would be responsible for defining and coordinating the development of the end-to-end software and data ecosystem.
The main goal of this model is to improve the interoperability of \gls{software} within the larger ecosystem by spreading responsibility for producing interacting pieces of \gls{software} to unique groups.
Simultaneously, this model would improve buy-in of the community for interoperability by each group having a stake in the outcome.
After initially scoping out the system, work would be divided into smaller work packages that could then be taken on or bid on by existing projects, institutes, and stakeholders to deliver pieces of the end-to-end system.
The new virtual institute would then either have its own funds to support successful bids, or would advise funding agencies on viable solutions to support.

The virtual institute would be a multi-site, federated organization that serves as a layer on top of the existing facilities. 
Each site would be responsible for one (or more) work-packages that address a component of the end-to-end system. 
All software created in the work packages would be deployed identically at all sites, and each work package provider must take into account all requirements and needs of all sites to be fully service-oriented. 
Examples of work packages could include archive interface, data models, \gls{VO} services, storage and data-transfer systems, science platforms, authentication and access control, pipeline infrastructure, \gls{community software}, and so on.
Regular audits would verify that, for each work package, the needs of all stakeholders are met. 
A board composed of representatives of all sites would decide about the work package splits and any evolution of the organization (i.e. definition of new work packages).

\subsubsection{Opportunities with this model}
\begin{enumerate}
\item Each site is a full expert in the subject matter they are responsible for.
\item User experience: The user experience is vastly improved. All users see the same interface at all sites. For many services, they do not even have to know where the data is located: every piece of data can be discovered everywhere. They also profit from single-sign-on and interoperable science platforms. The homogeneous user-experience will speed up the work of the astronomers and therefore increase the science output).
\item Localization: differences between the facilities and the ``branding'' of each site is achieved through localization inside the \gls{software}. In addition each site keeps their web presence.
\item Long term preservation could be improved by having identical software deployment mechanisms at different sites. This would make data migration easier in case a project ends or a site must close.

\end{enumerate}

\subsubsection{Risks of this model}
\begin{enumerate}
\item If each work package sites does one thing. staff may find themselves in stagnant career development situations (unless they change place of work).
\item If work package owners who now cannot control there own timelines or workflows for products, are dependent on another work package owner, there will be a lack of buy-in. 
\item If it is not carefully spelled out who is responsible for coordination, delays and prioritization of resources across the groups then it could be the case that there is no Owner of the risks, and the goals may not be met.    An owner of oversights of needs, interdependencies and timelines are also part of this risk that must be well managed.
\item Innovation could be stagnant if all parties are not ready, don’t have resources or don’t see the need to change at the same time. Creativity for new products likely requires changes for all work packages because of interdependencies. 
\item If there are no ``directors'' who can see and support the larger goals, these risks are all amplified.

\end{enumerate}


\subsubsection{Prototypes or examples}

The Gaia Data Processing Consortium (DPAC) is a collaborative effort involving several European institutions and organizations responsible for processing and analyzing the data collected by the European Space Agency's Gaia satellite. 
The consortium's organizational structure is comprised of several Coordination Units (CUs), each specializing in different aspects of data processing and analysis. 
These CUs are responsible for tasks such as data pre-processing, astrometry, photometry, spectroscopy, and more. 
Each CU is set up to receive data from a previous step in the processing and pass data on to subsequent CUs --- that is, the work is somewhat siloed.
DPAC has generated many successful data products and releases that demonstrate the efficacy of this model.
However, the Gaia DPAC is a more focused and smaller mission than developing a data and software ecosystem to serve all of ground-based astronomical research.
The siloed approach to a consortium can lead to communication and integration challenges between units, hindering effective collaboration and data consistency that will be critical for a scaled-up version of this model that would serve a broader community. 
Lack of cross-domain expertise and potential duplication of efforts are additional drawbacks that may impact such a consortium's efficiency and scalability.




%% file: conclusions.tex
\section{Conclusions} \label{sec:conclusions}

This report summarizes the outcome of a workshop conducted in February 2023,
led by the \gls{NSF} and the Simons Foundation, and attended
by representatives of many of the stakeholders in astronomical software and archives from the \gls{NSF}, NASA, and \gls{DOE} communities. The workshop was sparked in response to the recommendations in the Decadal Survey described in Section~\ref{sec:intro}.

We have described the challenges facing the astronomical community
given the volume and complexity of data in the modern era (Sections \ref{sec:problem}
and \ref{sec:ecosys}). 
The attendees of this workshop engaged in a week of reflection, discussion,
and argument regarding the nature of the challenges, their potential technical
solutions, the structural models that could implement those challenges, 
and many related issues. 


During that week, the group reached a consensus that a crucial requirement is a well-funded, efficiently coordinated, and sustained long-term initiative spearheaded by a new entity equipped with a clear mandate, adequate resources, and strong community support. Defining the nature of this entity remains a pivotal question. The participants at the meeting outlined the range of possibilities, roughly defined by the four structural models outlined in Section \ref{sec:casestudies}.


The workshop's primary recommendation is for the U.S. astronomical community to adopt the roadmap outlined in Section \ref{sec:scope} to determine the most appropriate structural model. This entails commencing with a Steering Committee backed by federal funding agencies, which will engage the broader astronomical community in formulating the scope, requirements, and organizational framework for the proposed entity. Given the time required for this development process, it is imperative that it kick-starts promptly.

The hurdles ahead are surmountable with the infusion of fresh resources wielded by an organization empowered to take action. The success of this endeavor would unleash immense potential that might otherwise go untapped, particularly in light of the extensive new observational datasets slated to emerge in the forthcoming decades.

%% file: acknowledgments.tex
\section{Acknowledgments}

We thank the Center for Computational Astrophysics (CCA) at the Flatiron Institute (a part of the Simons Foundation) and especially the administrative staff (Kristen Camputaro and Fatima Fall) for administrative support and organization for the ``Future of Astronomical Data Infrastructure'' workshop that provided the catalyst for this document.



%% file: dives.tex
\section{Deep Dives on Technical Components and Barriers} \label{sec:dives}

\input{opsplat}

\input{pipelines}
\input{dragons}
\input{crossarc}

\input{sciplat}

%% file: opsplat.tex
\subsection{Deep Dive: Proposal to Observation} \label{sec:opsplat}

\Lead{Bryan Miller}
\Contributors {Kim Gillies}

\subsubsection{Introduction}
Over the last couple of decades the development of queue-based observing models and telescope networks has made telescope operations much more dependent on complicated software systems.  While the traditional observational astronomy project workflow has always consisted of proposals, observing, analysis, and publication, the time on new, large ground-based facilities like Gemini and the VLT at the turn of the millennium was sufficiently rare and valuable that they adopted more queue observing made popular by the Hubble Space Telescope (\gls{HST}) and other space missions (\figref{fig:cycle}). Ground-based queue, or service, observing is like space observing in that the observatory carries out the observations based on scheduling constraints (e.g. weather) in order to make better use of the best conditions and allow more flexibility to respond to new, high-priority observations such as supernovae or \gls{GRB} follow-up and to manage programs with complicated timing constraints. This “cycle of discovery” workflow for observational experiments requires more tools for observation preparation since all queue observations must be prepared in advance and making both raw and processed data available in an archive so that the community can do more science with existing data.

Each node in the cycle of discovery and the connections between them are enabled by \gls{software} applications and interfaces. This chapter focuses on the \gls{software} needs for the first half of the cycle: from proposal preparation through observation execution. Other chapters in this report expound on the \gls{software} needs for handling, processing, analyzing, and archiving the data.

The first half of the cycle of discovery is important for helping science teams find the appropriate telescope capabilities for their experiments and propose for the observations that will be useful for achieving the scientific goals, and for enabling observatory staff to collect the data efficiently and with the expected properties. Software usability must be given thoughtful consideration in order to reduce frustration and encourage the use of the facilities. In order to maximize science observatory \gls{software} tools must aid users in achieving their goals. The type of tools used at each step of the process are summarized below.

Space missions and ground-based survey facilities follow a similar cycle of discovery but the first half of the cycle consists of project evaluation by funding agencies and then survey operations. 

\subsubsection{Capability Discovery}
\begin{figure}
\begin{centering}
\includegraphics[width=0.8\textwidth]{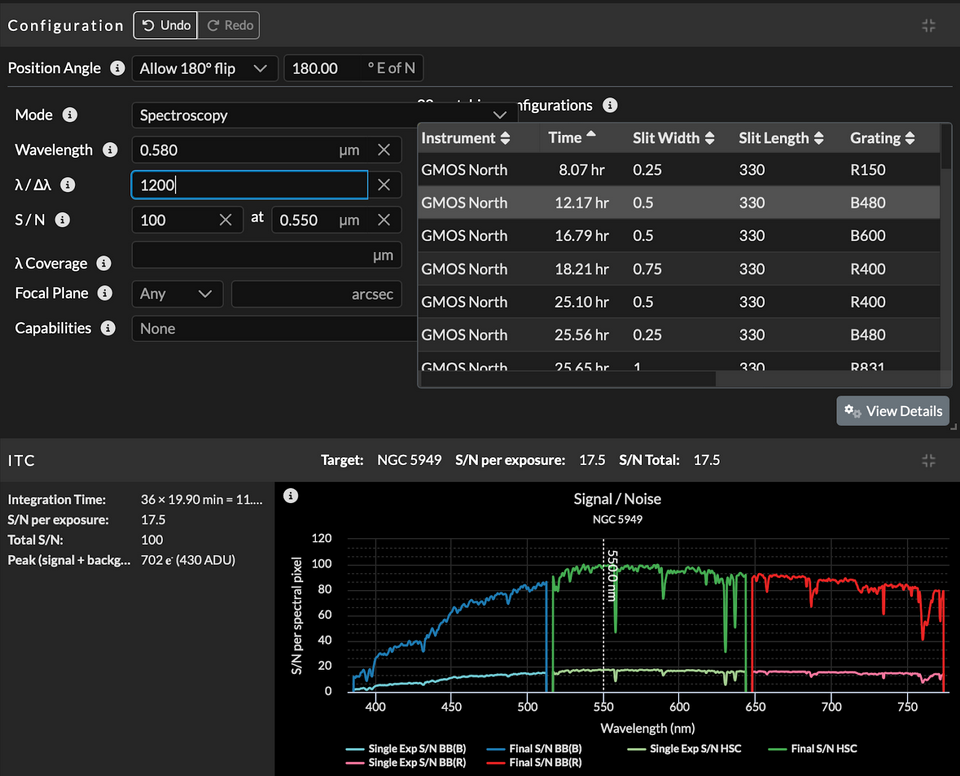 }
	\caption{
The “Phase 0” or capability discovery in Gemini’s Explore application that will be used for both proposal preparation and observation definition. A user enters their science requirements (mode, wavelength, signal-to-noise, etc) and they receive a list of capabilities that meet the requirements with the time needed.
\label{fig:apt}}
\end{centering}
\end{figure}

Once astronomers formulate a scientific question to be answered, they must determine whether any facility exists that can carry out the experiment. Historically this has been done by scouring observatory web pages, reading instrument manuals, and experimenting with integration/exposure time calculators. This process is inefficient and teams may miss feasible options. A new alternative being implemented in the \gls{GPP} Explore application is to include all of the Gemini instrument capabilities in a database and then give users a list of all instrument configurations that meet the science requirements such as wavelength coverage, resolving power, signal-to-noise, etc. (\figref{fig:apt}). Integration or exposure time calculators are integrated so that users can see how long each option will take and make a well-informed decision about which \gls{configuration} is best for their science.

\subsubsection{Proposal Preparation}
Observing proposals (Phase 1) are usually submitted once or twice a year for observations that will be done in the subsequent 6 or 12 months. Word processing software such as Latex, Word, and web-based equivalents are usually used to write the “essay” sections, such as the science justification and experimental design. PDF versions of these documents are usually included with the full proposal. Other information such as investigator contact information, target and instrument details, and time requests are usually required in a machine-readable format for subsequent stages so they are collected by a proposal software tool. This can be a standalone application like STScI’s \gls{APT}, \gls{ALMA}'s Observing Tool and Gemini’s Phase I Tool but many organizations are moving towards web-based systems. Integration or exposure time calculators are important tools in this process for calculating the observing time needed to reach the science goals. Integration of these tools can speed up the proposal preparation process.

\subsubsection{Proposal Evaluation}
Software tools used for managing proposal files and the TAC ranking process include web interfaces to databases, integration time calculators, and scheduling algorithms for filing queue time or assigning classical/visitor nights. Facilities with multiple partners or that allow joint time allocation with other observatories, e.g. NASA missions such as \gls{HST}, JWST, and Chandra with NOIRLab ground-based facilities, have to develop procedures to transfer TAC rankings and comments and target/instrument details.

\subsubsection{Observation Preparation}
Queue, or service, and survey observing modes require that observations be prepared well in advance of when they will be executed since they could be done whenever conditions are appropriate.  “Phase 2” \gls{software} tools help users define the observation details such as specific instrument configurations, offset patterns, timing constraints, and necessary calibrations. Integration time calculators are used again for signal-to-noise and exposure time calculations.  These tools have been standalone applications such as the APT and Gemini’s Observing Tool, but these are also moving to web-based systems. Finder charts must also be created to aid target acquisition and sometimes guide stars must be chosen. The observations are stored in database systems at the observatories for scheduling and execution.

Follow-up observations of fleeting time-domain events such as supernovae, \glspl{GRB}, fast-radio bursts, and gravitational wave and neutrino detections requires \gls{software} automation for faster, in the moment, observation definitions. Facilities need to provide \glspl{API} for communicating with external \gls{software}.  The community is then developing tools such as \glspl{TOM} (e.g. Las Cumbres' TOM Toolkit) that use these \glspl{API} to allow teams to import targets from \gls{TDA}/\gls{MMA} alert “streams”, submit observations to facilities on which they have approved time, and download the resulting data from archives.  These tools are also useful for managing projects with lots of targets or large teams.

\subsubsection{Observation Scheduling}
Traditional “classical” or “visitor” nights, during which the proposing teams take the data, must be scheduled when the targets are visible and the moon phase is as requested. During queue or service observing, observations must be executed when conditions or time constraints are met. This is made more efficient with software tools for planning and observation selection. Robotic telescope networks designed in large part for time-domain follow-up require sophisticated automatic scheduling algorithms (e.g. Las Cumbres network, \gls{AEON}, Skynet, \gls{ASAS-SN} ). These algorithms are also used by space missions and for improving efficiency and response times for manually-operated ground-based telescopes.

\subsubsection{Observation Execution and Data Collection}
Modern telescopes are basically computer networks with real-time systems working in coordination. Robotic and survey telescopes are fully or nearly automated. Manually-operated facilities require \gls{software} interfaces for the users to control and monitor the systems. Observers must view any scheduler output and then execute the observations. Once this starts, users must interact with the data processing systems mentioned in the other sections of this document. Rapid data processing and visualization are needed for target acquisition and initial quality assessment of the data. Any data that does not meet requirements must be re-observed.
Requirements are part of the observing proposal and may include  physical parameters (sensitivity, resolution, source extension) to be achieved.
The \gls{software} that executes the observations also works with the instrument \gls{software} to write the header meta-data that is essential for reducing the data.

\subsubsection{User Support}
Science teams need to be supported by observatory staff throughout the cycle of discovery.  Commercial platforms such as \gls{JIRA} are typically used for collecting and tracking support requests.

\subsubsection{Limitations}
If the cycle of discovery represents the workflow of so many observatories, why aren’t there common tools shared by observatories? There have been successes in the past. Gemini and Keck, for instance, both used the \gls{EPICS} as the basis of their control system, which continues to thrive to this day as a project with broad support. Also, the Large Binocular Telescope adapted Gemini’s Phase I Tool and Observing Tool when they implemented a queue observing mode. The biggest limitation is that observatory \gls{software} has been tied to specific projects, each with its own requirements, cost, and schedule. Maintaining construction deadlines can lead to very siloed approaches.

New large facilities occur infrequently and have long lifespans. But \gls{software} tools, environments, and best practices change at a very fast rate. It is a difficult decision to reuse a \gls{software} tool for a new project that may be based on technology that is already dated even if it has been successful. Ten or twenty years ago creating a complex user interface in a web browser was risky or impossible, but that is now the obvious technical choice, and creating applications that must be installed is out of the question.

Because \gls{software} technologies change so rapidly it is difficult for existing facilities to keep their \gls{software} current. Older \gls{software} becomes harder to maintain. If it is not updated regularly, then the effort to refresh it becomes much larger and more expensive. More resources or knowledge in the community about new technologies would encourage facilities to keep their \gls{software} updated. This would then make it easier for new projects to reuse or build on existing \gls{software} or frameworks.

The cycle of discovery is a shared, common workflow, but observatory facilities and operations are not the same; each has its own set of unique capabilities and requirements. The \gls{software} is where these differences are most visible. Some \gls{software} areas must be specific to the facility, but in other areas there are opportunities for commonality and potentially reuse. A \gls{software} tool that is shared must separate what is common from what is specific to an observatory. There are also limitations in existing \gls{software}. Desirable capabilities such as cross-observatory proposals are not widely available, because the \gls{software} tools are not shared and do not communicate programmatically.  It is only recently that \gls{software} technology has evolved such that imagining reusable observatory \gls{software} infrastructure that can account for these issues is possible.

\subsubsection{Commonalities}
The process of developing the operations \gls{software} shares many issues with the development of data management \gls{software} besides shared development tools and practices. Hiring and retaining qualified engineers is a challenge since salaries in astronomy tend to be lower than those in industry. Because of the skills required, the engineers needed to build the operations \gls{software} tend to come from computer science backgrounds, rather than astronomy. Therefore, astronomy facilities are competing directly with industry giants such as Google, Facebook, and Twitter, for talent. Fortunately, astronomy has many interesting technical as well as scientific problems to solve, which attracts people to work on them.

The groups that develop operations \gls{software} should also be encouraged to make all code publicly available and to upstream changes to third-party \gls{software} that is used. Besides making more sharing possible, many developers get satisfaction from contributing to generally used \gls{software} libraries.

Operations and data reduction software also share many tools and services. Currently both require web application interfaces and databases. Both make use of the same astronomical catalogs and queries (e.g. Gaia for guide stars and astrometric and photometric calibration) and the same calculations for nightly ephemeris data and object positions. \gls{Operations} and data management software, which work together directly during data collection, are both required to produce the best data for science experiments.

\subsubsection{Opportunities}

Despite the limitations and technical challenges, the \gls{software} described in this section presents a unique opportunity for the creation of a system of \gls{software} tools that could be reused by future, new observatories and that could also be integrated into the operations of existing large and small observatories. The advantages of commonality to telescope users and staff would allow more time to focus on science outcomes and less on learning new tools. From the institutional perspective, commonality and reuse should result in cost savings in the long-term.

In our view, a common approach to support proposal preparation through observation execution systems needs a dedicated staff, with requirements, schedule, and deliverables. Similar to AstroPy, the project would bring together a collaboration of contributors with interests and experience in the different phases and their tools and technologies. Some project deliverables could start with existing tools and generalize them, and others may require the creation of new tools. The previously mentioned EPICS project is a great example of what is needed to create a broadly useful \gls{software} system that is maintained over the long term. A key to success is that the effort produces deliverables quickly and iteratively in order to show progress and gain feedback from stakeholders that can improve subsequent releases. The following are possible areas that could be included:

\textbf{Capability Discovery and Evaluation} – the creation of protocols and services that allow astronomers to compare and determine what facilities are available for a science objective is an area that can reduce the time to create a science proposal. At a higher level than integration time calculators or instrument simulations, these services take a description of the science and return an observation \gls{configuration} that can accomplish the science, allowing the potential investigator to easily decide if a proposal is worthwhile.

\textbf{Proposal Preparation and Tracking} – Quite a bit of the information that goes into an observatory science proposal is common to all proposals. A data model for proposals that includes the common data but also unique information could form the basis for a set of services that support proposal preparation and submission. There is quite a bit of commonality in the processes associated with evaluating proposals and with the shared model for proposals, there is an opportunity to create \gls{software} that could be used by a variety of observatories.

\textbf{Observation Planning and Scheduling} – Observation planning is the most observatory specific of the software tools an observatory uses in the cycle of discovery. A common data model for observations and backend services for handling storage, updates, and queries of observations and science programs is one opportunity to add support for creating reusable tools in this area. A common data model allows the creation of common scheduling tools that implement an observatory-specific policy. Shared expertise on scheduling algorithms and scheduling ``toolkits'' (e.g. Las Cumbres’ \gls{OCS}) would make these easier for new facilities to implement.

\textbf{Data Pipelines at the Telescope} – Data processing at the facility is a critical part of observing. Data from detectors must be processed during target acquisition to ensure the telescope is pointed correctly and again following the observation to make sure the data meets the requirements of the science program. Increased commonality in processing \gls{software} and pipelines can help at the telescope, too.

\textbf{Data Transfer and Submission} – a package of software that allows the synchronization of data collections between sites and efficient transfer of data between sites is a tool area that would provide value to large and small observatories. This software would make it easier for smaller observatories to be part of the end-to-end system, and make their data more valuable and available to a wider audience. Creation of standards or recommendations for \gls{metadata} and file definitions would be valuable to enable integration with upstream archives.

The differences between observatory operations procedures, telescope and instrument idiosyncrasies, and traditions create challenges for the goal of providing shared software infrastructure to support the telescope part of the cycle of discovery. The previously suggested software tools focus on protocols, data models, and services and not user interfaces, because the user interface must contend with the differences between observatories. It’s possible that common user interfaces for some of these tools are possible if there is agreement on data models and common requirements in the areas such as workflow, \gls{UI} design and behavior, visualization, and deployment.

At the technical level of \gls{software} development, all \gls{software} produced in this area must be open source, available, and come with licenses that do not restrict usage. Common authentication mechanisms should be encouraged that allow integration of tools in a seamless, standardized way that makes web access as easy as possible. The \gls{software} needs to be constructed to allow updates and evolution over a long period of time to follow changes in computer and programming technologies, user expectations, and new ways of observing.

These recommended changes would help existing facilities in the middle of upgrades and with supporting other 2020 Decadal Survey priorities of \gls{TDA}/\gls{MMA} follow-up networks and future extremely large telescopes (e.g. \gls{US}-\gls{ELTP}).

%% file: pipelines.tex
\subsection{Deep Dive: Pipelines} \label{sec:pipelines}

\Lead{ Don Neill}
\Contributors{William Vacca, James Turner, Kathleen Labrie}

\subsubsection{Introduction to Pipelines}

We define a pipeline to be the software modules and associated infrastructure that takes raw data (or simulations) as input and processes that data to generate science-ready (and archive ready) data products.  Each project defines what is meant by ``science-ready'' and  that can range, as example, from an instrument-artifact corrected and calibrated image, to redshift or other extracted parameters.  The components of a pipeline typically consist of data reduction and analysis tools, automation infrastructure, user interfaces, logging and reprocessing capabilities, product ingest to a data archive, and ancillary tool kits.

In our experience, pipelines provide a very concrete example of the potential for code re-use and community-based augmentation as most space and ground-based telescopes support data processing pipelines for data reduction.  There is overlap with common scientific algorithms used in data reductions among projects and automated infrastructure that could be shared as long as scale of data \& scope of processing align.   Some existing pipelines have ``community-contributed'' software reduction and analysis tools associated with them.  Potential  improvements include having a standard for incorporating community contributed software into a project pipeline, and modular organization that encourages uptake of  a pipeline framework by new projects.

\subsubsection{The Problem}

Currently, data processing pipelines are the locus of much duplicated algorithmic software.  The potential exists to have a common infrastructure for automation but this is seldom done except within institutions.  Libraries, such as AstroPy, offer the opportunity for code sharing but further community contribution and uptake in the area of pipeline processing needs to be fostered.

The FITS standard imposes some limitations on pipeline data products, especially those pipelines dealing with complex reductions where multiple calibration sources need to be combined in order to remove instrumental signatures.  The FITS standard has been widely discussed in various forums and several new file formats have been identified that combine FITS with other hierarchical file structures.

The priority of pipeline software and documentation within instrument development budgets presents another problem.  Often ground-based instruments are delivered without fully developed and well documented pipelines since operational software is considered low-hanging fruit for budget cuts.  

The methods used to acquire calibrations impact much of the pipeline infrastructure.  Associating calibration sets with science data is one of the primary functions of a data reduction pipeline.  Currently there isn't a good community forum in ground-based astronomy for exchanging ideas about calibration strategies and best practices for forming associations.

The way that a pipeline records the steps that have occurred during processing is currently determined by each project.  No standard exists for keeping track of what reduction or analysis steps have been performed on a given science product output by a pipeline.  At ESO, the Reflex framework for pipeline workflows provides such capabilities and needs a closer review.  Definitions for what belongs in reduced image headers are also currently not standardized, though there is an IVOA standard that defines a subset of keywords to enable data discovery and that could be used as a baseline for further standardization.

Given the durable nature of digital data, it is important that pipeline software have lifetimes commensurate with the data.  Pipeline software that can be used to re-reduce raw data or extract new parameters from
reduced data needs to be curated as a value-added product and associated with the archived data products.

Another area where work is needed is to standardize what useful intermediate pipeline products need to be archived along with the raw and reduced data.

\subsubsection{A vision for change}

The vision stated herein can initially be addressed by providing a community forum for the exchange of ideas.  As envisioned, the forum supports an organic development of community standards that will promote good documentation, code re-use, interoperability of pipeline modules, and output data format standardization with all the concomitant benefits.  We believe community engagement will benefit pipeline developers, as well as end user scientists and data archivists.

We can summarize our vision as follows:
\begin{itemize}
    \item Adopt pipeline development best practices that promote code re-use.  These practices include: making pipeline software well-designed and modular for multi-project use;  clearly documented modules; adopting a coding style that allows code to be clearly read; and use of a publicly visible configuration management repository.
    \item  Pipelines are  extensible such that new recipes can be built from the existing pipeline modules.
    \item Define and adopt a standard for incorporating community generated  algorithms or modules into pipelines.  A similar standard for adding pipeline modules to upstream libraries should also be discussed and defined.  These standards must include criteria for scientific quality, coding style, unit testing, and documentation.
    \item  Build a close relationship between pipeline developers and archives as most data, raw and reduced, are destined for a data archive.  In particular, header keyword standards must be developed with input from archives for repositories to associate/store data products and with project/community scientists for metadata required for data analysis.
\item Augmentation of the FITS standard needs continued discussion.  Current tools for handling hierarchical data within headers needs further examination and the definition of a best practices generated.  A standard set of tools for reading and writing data and headers should be developed, or extensions to existing libraries to cover more use cases. 
\item The range of pipeline interfaces must be considered, from batch to highly interactive as is appropriate for the specific instrument.  Operations typically benefits from batch processed pipeline data for efficiency, while observers may require atomic execution of pipeline steps to review data and set parameters.
\item All instrumentation projects provide specific line-items for data reduction and software in their budgets such that software budgets are protected and not used as contingency for hardware overruns.
\item  Define standards so pipeline developers know how to work with the open source pipeline repositories and the framework that is developed from these efforts.    
\end{itemize}

A pipeline development model includes modular infrastructure components, automation features, common calibration and data reduction modules, standardized data product and header definitions, and logging and reprocessing mechanisms compatible with data archives and science needs.  With the right coordination and funding model, we envision an open source approach where projects with a common base work together to build shareable software to achieve these goals.

%% file: dragons.tex
\subsection{What We Learned Raising DRAGONS}
by Kathleen Labrie.

Gemini Observatory has the mandate to provide data reduction software to its users’ community for all its facility instruments.   As our initial suite of tools grew to support more instruments, the code duplications and design inefficiencies multiplied to the point that development and maintenance outstripped our resources.

We needed a new model aimed to ensure sustainability.  The core principles we adopted were:
\begin{itemize}

\item A uniform internal representation of the data, headers, and file structure.  This opens up code reuse on a large scale.
\item Invest in generic algorithms.  Take the time to think and write algorithms that can be reused across instruments and modes.
\item Implement automation based on the data, not on information provided by the user.  This includes automated calibration association.
\item Still allow for user customizations and optimizations, as desired or required to achieve specific scientific objectives.  This includes interactive tools.
\end{itemize}

Our new model led us to a multi-purpose pipeline infrastructure that can automatically reduce data from multiple different instruments and modes (imaging, various types of spectroscopy, optical, near-infrared) using libraries of reusable algorithms.  Our implementation of the new model is DRAGONS – Data Reduction for Astronomy from Gemini Observatory North and South.  The unique framework, coding standard, coding style and overall philosophy of DRAGONS have proven to be a lot more cost effective than our previous platform.  The team developing and carrying out the long-term maintenance of the code base needs to learn only a single system, not a multitude of bespoke pipelines.  Knowledge transfer between the team members is also facilitated.  However, in this model the initial development phase requires additional time and this pushes against the internal and external pressure to produce data reduction code “immediately”.  A balance between sustainability and speed can be hard to achieve.

Developing a new platform while maintaining the old one is also an important challenge.  Currently, we do the minimum needed to keep the old platform (Gemini IRAF) alive as we transition support for each instrument and mode to DRAGONS.  

Hiring people with the right skills is difficult.  Fewer people learn how to reduce raw astronomical data nowadays compared to two decades ago and those who do are not always savvy programmers.  Hiring is costly, time intensive, and drains existing resources to train the new hires, making retention critical.  Frequent turnover also affects the team dynamic. 

A clear avenue for the growth of DRAGONS is through streamlining usage and development of third-party DRAGONS packages.  While the technical solutions are fairly simple, the big hurdle is the documentation, or rather the lack thereof.  Documentation is key to sustainability yet it is always the first thing cut when a project is under pressure. 

%% file: crossarc.tex
\subsection{Deep Dive:  Long-term archival \& Cross-archive Collaboration} \label{sec:crossarc}

\Lead{Tess Jaffe, Mario Juric}
\Contributors{Rafael Martinez-Galarza, Stephen Bailey, Gus Muench, Benjamin A. Weaver, Anna L.H. Hughes, Rob Seaman}

\subsubsection{Prime directive 1:  keep all the bits}

Long-term preservation of data is not a new topic, but it remains to be properly addressed in astronomy in general.  Individual solutions exist in different domains, but they do not cover all of the data produced by taxpayer funded projects in the \gls{US} (much less beyond;  but the \gls{US} is the scope of this document).  Outside of astronomy, there are organizations that have experience curating knowledge over a millennium such as the Vatican and various universities. Arguably, 50 years in astronomy is an equivalently impressive time scale given the pace of change in the storage formats and media used.  Likewise, there are examples of government funded data centers that have curated such data for decades. But there are still gaps. Some agencies such as \gls{NASA} are updating funding policies to require that the results of all funded research are preserved, but it is not always clear what is the appropriate repository for the results (e.g., does Zenodo count?).  Each agency needs to clearly define long-term preservation requirements (e.g., \gls{FAIR}) and a funding strategy to meet them. All projects, large and small, should know where they can go to deliver their data for archiving, including everything from raw detector outputs to high level science-ready products.  And all projects should be required to do this following the archive’s standards and procedures and with the archive’s help. It is not necessary that there be one archive containing all data, only that all data end up in at least one such qualified archive so that no bits are ever lost. Once this minimum is established, then we can move on to the question of ensuring that the data remain useful as technologies evolve and that the archives collaborate and interoperate rather than remain siloed.

\subsubsection{Prime directive 2:  keep all the software}

Along with preserving the actual data bits, effort is needed to preserve software necessary to interpret the data, along with the ability to execute it in the future.  The community can and does support activities to preserve analysis software as well as to maintain packages that access standard data formats, e.g. astropy.io.fits. This support should continue. However there is a risk of losing the capability to discover and run software that interprets data stored in specialized formats.  For example, SDSS provides “Atlas” images, which are very small “postage-stamp” cutouts of the pixels around a specific photometric object. Although these images are stored in a FITS file, simply opening the file with astropy.io.fits is not useful, because the actual images are stored in a specialized binary format. Instead a stand-alone C program, read\_atlas\_images, is provided to interpret the binary data and extract the requested images. Similarly, point-spread function (\gls{PSF}) data may be stored in a specialized format (the same SDSS read\_atlas\_images package also interprets \gls{PSF} data). Other examples include image masks and, in spectroscopy, resolution matrices.  Finally an example is IRAF, that is still widely used, and is no longer natively executable on modern popular hardware (e.g., 64bit architectures, or Apple's ARM-based chips).

\subsubsection{Prime directive 3:  keep it discoverable}

There are many levels of data documentation that need to be preserved. Many projects publish papers describing their data, but these do not necessarily capture all knowledge about the data that may be known to project participants, in particular, there may be details about data files being preserved that may not be fully described in published papers. Every aspect of documentation that is needed to interpret the data correctly should be preserved.  In addition to published papers, many projects provide high-level documentation websites with many details about data structure, access, instrumentation details and so on (e.g. \gls{SDSS}). These websites need to be preserved in some form after the formal end of the project. At a lower level, every preserved data file should be described in some form, for example a \gls{FITS} file should have a human-readable description that describes its contents: what each section (\gls{HDU}) contains, what headers to expect, the definition and units for table columns, and so on. \gls{SDSS} has an extensive example. Finally, it may be necessary to preserve and mine internal correspondence for knowledge about the data, in case the higher-level documentation is incomplete in any way. This could be archived email lists, help forums, Slack channels and similar communications. While the lower-level resources may not be frequently accessed by the general astronomical community, they are essential to data curation and would be used frequently by archivists.  They are also important resources for any “help desk” that is actively answering questions from the public.

Not all data will remain equally relevant for the community.  New observations with improved instruments may supersede previous data, or they may be superseded by new calibrations or processing of the same dataset.  Similarly, higher level final data products will likely remain in active use longer than lower level raw data or intermediate products.  Pragmatic considerations of storage resources, budget, and human curation effort may require moving some datasets to “cold storage” for long term preservation even if they aren’t rapidly accessible.

At the same time, it is critical that these archived datasets remain discoverable and retrievable.  A seminal paper may use a dataset that has been superseded by a more recent one, but the original dataset is still needed for understanding reproducibility details of an analysis, especially if a newer analysis on newer data finds a conflicting result.  Time domain studies benefit from historical data, even if there is more recent “better” data for the same objects.  And given the broad diversity of ideas coming from the community, a dataset that is not currently in active use may become useful again for a new idea in the future.  Data discovery tools should continue to provide access to the \gls{metadata} about these archived datasets, while also providing indications if there is a “better” recommended dataset for most users, i.e. the discoverability of older archived datasets should not come at the expense of the discoverability of current datasets which could be “lost in the noise” if all datasets are presented on equal footing.

\subsubsection{Prime directive 4: upgrade when needed}

Similarly, there will always be costs and benefits to be weighed when it comes to maintaining or upgrading from old data formats to new and from old data-interpretation \gls{software} to new.  Data is not accessible if it is not readable or interpretable.  But how easy the data is to read or interpret (how accessible) should be a matter of considered priorities, needs, and resources.  Where possible, priority should be placed (and funding should be allocated) on developing generalized tools and documentation that can be used for translating from old formats to new as needed (note that this doesn’t just apply to hard data-format type, but also to other types of informational formatting, such as meta-data standards, which over time may evolve and shift).  Thereby, old, (especially popular) formats should be supported for the community as a whole as long as is practicable.  That will allow archives to continue to store data in old formats if they do not have the resources to upgrade, and will also allow those in the community to retain accessibility to personal data sets that may not be otherwise archived.

But this is not to advocate the extreme scenario of archives never updating the formats of the data they are storing.  Rather, efforts to update datasets should be done as targeted projects taking into account the needs of both the archive and the community.  If an older dataset has recently become newly popular it may make sense to improve its accessibility by updating its format.  Or, if an archive believes it will likely lose capability to support an old format then the dataset may need to be upgraded to prevent its loss.  By monitoring data usage and storage needs, archives can be in a key position to flag what new tools and efforts may be of best benefit both for the archive and for the community.

\subsubsection{Parameters of solutions}

When trying to design lasting solutions for Astronomy data storage and access, it is important to keep in mind where the costs really lie.  Because we are entering an era of enormous data volumes, and because we still tend to imagine any ‘product’ like it is a physical thing, most who work outside data-centers latch onto the equivalent physical thing - data-storage space or hardware - as representing the bulk of the cost.  But this is entirely wrong.  Not only is the purely physical aspect of any digital data-set completely worthless, but its scientific value is only, sometimes, tangentially related to its volume.  ALL of the bits are precious, because almost none of them, if lost, can ever truly be replaced.  And NONE of the bits are usable, at all, if we don’t know how to read and interpret them.  And maintaining that knowledge and ability, of how to use, access, and interpret the data, is the expensive part.

Maintaining knowledge and accessibility to data costs labor hours.  Unlike data volume, estimating and quantifying that cost is difficult and complicated, and usually comes with trade-offs (e.g., maintenance of historical data  vs. access to current solar observation data taken today).  The risk is that we may lose precious data if a long-term approach to handling these questions are not addressed.  Maintaining knowledge and data accessibility means being able to read in the format the data is stored in.  Formats age, and we need either campaigns to transition data to new formats, or we need to actually build tools that will allow for translation when needed.  Software tools made for reading and analyzing a dataset become dated, their base-code maybe even deprecated, and not only labor hours, but also the right knowledge base is  necessary to replace those tools.  Maintaining data accessibility means knowing enough about the data to be able to use it. Complete documentation is required, that doesn’t leave out the details of what “everybody knows” at the time the data were generated.  And while estimating labor hours for proper documentation is hard, one thing that is always true is that it gets more and more expensive with time.  People forget; the relevant notes become lost; a project ends before the task is complete.  Saving the bits and being able to read the bits isn’t even enough if people aren’t able to discover the data.  Labor hours are also needed to properly catalog and curate data, and then to make sure those catalogs and curation tools are themselves well-enough maintained to continue to be usable.  These labor-hour costs depend on the complexity, nuances, and relative age of the data (and the data archive).

Funding agencies should define and fund long-term archives and require that funded research projects make results available there. The NASA astrophysics archives have explicit mandates to archive NASA-funded project data in perpetuity. These archives must also define metadata standards and provide user-friendly services to the community to help researchers submit their results for archiving. See, for example, the \gls{IRSA} web page on how to submit data for archiving.  See also \gls{NASA}’s new Science Information Policy, which expands the requirements to include not only mission data but also the results of individual funded research projects.  The existence of such long-term data preservation centers and their scientific curation staff is a prerequisite to much of what we recommend here.

Funding agencies should also establish one or more regular sources of funding for modernizing archival datasets and \gls{software}.  As described above, once a long-term archiving solution has been implemented, the job is merely starting.  After ensuring that the bits are kept in perpetuity, that the bits are well described by documentation and metadata, that the bits are formatted such that they can be read and processed by existing \gls{software}, then the task becomes to keep up with a changing technological context.  As also described above, long term data archivists are in a good position to know what datasets and \gls{software} need to be upgraded and how.  What the community, and in particular the agencies, need to do is to recognize that this will be a continuing cost and to provide funding opportunities to meet the needs of data modernization.  Some efforts have already been made to reconfigure certain datasets and \gls{software} to make them more useful in the modern computing context. The current interest in cloud computing is also spurring an interest specifically in adapting to one or more commercial clouds. But this is generally decided on an ad hoc basis and only when the need is severe enough to justify stretching already allocated resources. The existence of regular sources of funding for modernizing archival datasets and \gls{software} would allow this to happen more regularly and carefully.  A competitive funding call could convene a panel to review each project’s justification for the upgrade expense and compare it to others.

\subsubsection{Federation}
As elsewhere in this document, we want to emphasize a \textit{federated} approach to addressing the challenges of curating, archiving, and accessing astronomical data.  There is a broad diversity of user needs within the astronomy community, and that necessitates a broad range of solutions that naturally will be different from each other.  The needs of a single investigator wanting all available data for a single astronomical object are very different from the needs of a large survey collaboration applying High Performance Computing resources to jointly analyze the combination of two large datasets from different archive centers.  A single centralized approach simply would not be able to meet the needs of all users.  Some user needs are already well served, e.g. by the interoperability of the NASA archive centers, and the data curation at NSF’s NOIRLab \gls{Astro Data Lab}.  As we seek to expand the interoperability of these centers with new computing resources and datasets, it is important to not impose solutions that would break what already works for these communities.  While it may be perceived as redundant or wasteful if more than one solution provides access to the same data, this can be well justified if the different solutions provide different optimizations for differing communities and use cases.

%

%% file: sciplat.tex
\subsection{Deep Dive: Science Platforms}\label{sec:sciplat}

\Lead{Robert Nikutta <\href{mailto:robert.nikutta@noirlab.edu}{robert.nikutta@noirlab.edu}>}
\Contributors{Gregory Dubois-Felsmann, Ignacio Toledo, James Turner, John Swinbank, Julie McEnery, Rafael Martinez Galarza, Stephanie Juneau, Vandana Desai}

\subsubsection{ Science platforms today}

In the past decade, science platforms have emerged as a solution
to the challenge of efficiently mining the massive amount of
astronomical data generated by modern facilities. Traditional
computer systems struggle to keep up with the data processing,
storage, and network requirements. For example, surveys like
the Dark Energy Survey (DES) produced hundreds of terabytes of image
data and catalogs, and future surveys like \gls{LSST} will generate
several petabytes of data. The complexity of data and data products
has increased with the sophistication of instrumentation.

Modern data include for example, raw CCD data, reduced images, massive catalogs, multi-dimensional spectra, complex time series, heterogeneous file collections, derived high-level data products, engineering and operational datasets.  There are codes for data analysis and pipelines that are required to co-exist with the data  they were written for.  Science platforms provide a way to handle all of these data and the software tools needed to process them.

Data archives in the past allowed web queries to find and retrieve specific datasets, either through forms or more general SQL interfaces, but with limited user-facing compute capabilities. The big data avalanche has led to the advent of user-facing, general purpose compute resources co-located with the archives.
These science platforms promise easy access to massive datasets, powerful services connecting various data products, and substantial computing resources to perform data discovery and analysis tasks in physical proximity to the big data holdings, with support for both synchronous and asynchronous tasks.

More recently, the ever-increasing affordability and standardization of commercial cloud systems (e.g., \gls{GCP}, \gls{AWS}, Azure) has led to an infrastructure shift away from local data archives and
towards the \gls{cloud}.

In our view, the main drivers for the development of science platforms are thus:
data are too large and too complex for local storage and processing;
compute requirements for analysis are too costly for individual
researchers; \gls{software} installation and deployment must be
stable and versioned (reproducible); several large datasets need
to be combined for analysis (possibly across multiple physical
locations); collaboration with colleagues is desired (including
iterative analysis / development of algorithms); access to these
resources from different clients is needed (e.g., for
teaching purposes); standing data services are necessary for rapid
response to time domain events.

\subsubsection{User communities}

Science platforms target various user communities, often with
significant overlap. Some science platforms are built for the user
community of a particular survey (e.g., DESAccess for the Dark
Energy Survey, \gls{RSP} for Rubin's \gls{LSST}). In other cases they
are designed to support researchers interested in datasets from
the telescopes and missions operated by an observatory (e.g.,
\gls{NOIRLab}'s \gls{Astro Data Lab}, MAST's TIKE platform, 
ESA Data Labs, ESAP at the European ESCAPE project). Some have hosted multiple data archives (e.g., SciServer has SDSS and HEASARC data in addition to data from other disciplines like genomics) and/or have been deployed for multiple projects (e.g., additional SciServer instances have been deployed for eRosita and non-astronomical projects. In other cases, a single platform is built to support the astronomical community of an entire country (e.g.,  Brazil's LIneA and China-VO). Finally, platforms also exist that target general scientific communities within and outside of astronomy (e.g., SciServer, CyVerse, Canada's CANFAR).

Among any science platform user community, the researchers'
backgrounds and primary motivations range from individual researchers, to data scientists, to teachers and lecturers, students of all levels, and civic/grassroots scientists. In some cases observatory scientists and engineers use platform frameworks to improve operations, development and testing (e.g. the Dataiku-based engineering platform at ALMA).

Researchers at all career levels utilize open or semi-open science
platforms for a wide range of activities. The activities range from teaching astronomy and data science to student classes, participatory science projects
with hundreds of civic researchers, individual astronomers performing
custom data analysis, and entire science collaborations utilizing
a science platform's ecosystem of data, compute, and software.

The ease of use, the open (or semi-open) access to truly enormous data
holdings, the data services that enable curiosity-driven discovery,
and the provisioning of remote compute resources are an incredibly
powerful combination, drawing thousands of scientists and students to
science platforms today (e.g. Astro Data Lab has registered
\textasciitilde 2,800 user accounts since mid-2017; Rubin expects
\textasciitilde 5,000 users over the 10-year duration of \gls{LSST};
China-VO, if implemented as envisioned, will serve over
10,000 astronomy researchers). In addition, the delegated installation and
maintenance of common science and astronomy software packages at science
platforms liberates researchers from the burdens of downloading \& managing software installations.

\subsubsection{Common functional components}

Common functional components of science platforms include:
\begin{itemize}
    \item Authenticate and authorize users on the system.
    \item Host large collections of images and catalogs (maybe with different
    scientific focus).
    \item Provide database (DB) access to catalogs via the Virtual Observatory's
    Table Access Protocol (TAP).
    \item Allow DB queries through a webform and/or through custom APIs.
    \item Enable image discovery through their meta-data, usually via VOs' Simple
    Image Access protocol (SIA), and implement an image cutout service.
    \item Expose compute resources through a notebook interface (Jupyter /
    JupyterLab).
    \item Feature a comprehensive installation of any software packages needed
    by users (e.g., canonical Python sci/num/viz stack, astronomical and
    machine learning libraries, etc.)
    \item Give users visualization capabilities through general-purpose Python
    plotting libraries (e.g., matplotlib, plotly, bokeh), web-based graphics
    applications (e.g., Aladin Lite, pyWWT), and custom software (e.g., Firefly,
    CARTA).
    \item Provide data upload and download functions.
    \item Perform positional cross-matching (between in-house and/or
    user-supplied tables).
    \item Offer general and individual user support (documentation,
    workshops, helpdesk).
\end{itemize}
Depending on the field of specialization, scientific use cases, and the
available level of development effort, individual science platforms
may implement additional features, including:
\begin{itemize}
    \item Discovery and retrieval of spectroscopic datasets.
    \item Remote user file storage (often through VOSpace or WebDAV).
    \item Access to user-owned database space to store tabular data.
    \item Collaborative user groups / projects, with access management
    either by SP staff or by users themselves.
    \item Publication service for user-derived high-level science products.
    \item Direct or federated access to batch compute / HPC resources.
    \item Defining and running bulk data reduction pipelines (either at
    survey level, or by PIs).
    \item Integration with time domain resources (alert brokers, follow-up
    frameworks such as the Alert Event Observatory Network (AEON).
    \item Machine learning / model training capabilities (only CPU-based
    available so far, useful for teaching Machine Learning courses).
    \item Ability to execute legacy applications (for example containerized)
\end{itemize}

\subsubsection{Technical differences, and repeated implementations}

Most science platforms offer similar functional data services, which means
they can solve many common problems in similar ways. However, the actual
implementation of these solutions is often very different and unique to each
platform. Even when standardized protocols are used across multiple platforms
(such as TAP), the specific implementations are usually custom-made or
heavily modified versions of existing implementations, often without a
practical option to upstream custom development back to the root source.

Technology choices for science platform components also differ often.
In every individual case the choices are driven by specific requirements
of a platform or survey, familiarity of decision makers with a technology,
and funding levels (which often preclude commercial off-the-shelf
solutions).  Technical decisions to be made include networking file systems
(e.g., GPFS, NFS), data management systems (e.g., \gls{Rucio}, Butler),
operating systems (usually some choice of a Linux distribution),
database back-ends (e.g., Postgres, SQLServer, \gls{Qserv}, Oracle,
Greenplum), and the notebook compute environment (usually Jupyter
Hub, with Jupyter Notebook or JupyterLab as the front-end for
Python; but also Pluto for Julia).

In virtually all cases such decisions are made within the narrow
scope of the single system being developed. Most (and possibly all) science platforms
have not planned to be part of a larger ecosystem. Sharing of solutions
between science platforms is therefore complicated or impossible on a technical
level, even before ``the color of money'' is considered. Thus, science
platforms exist simultaneously, and often funded by the same agency,
which have all re-implemented solutions to common problems. There are
multiple cross-matching services, multiple TAP implementations, multiple
custom APIs to access databases, multiple visualization frameworks, etc.,
in existence.

Even when identical components have been selected (e.g., JupyterLab),
their customization can be incompatible with other science platforms.
That is, it may  be impossible to execute one science platform's Jupyter
notebook instance on another science platform, even if both have
containerized this service. How the user-facing software stack is
exposed and maintained, and whether users are enabled to install their
own \gls{software}, differs widely between science platforms (often by
virtue of policy).  There are many deployment modes for notebook environments,
including on-premises, at Google Collab, Binder, and various custom
deployments on commercial cloud providers.

\subsubsection{Barriers to interoperability and challenges to synergistic development}

Several problems from different domains currently inhibit
\gls{interoperability} between science platforms, and hinder synergistic
development and community maintenance of common science platform components.
Among them are:
\begin{itemize}
    \item In practice, only TAP and SIA seem to interoperate well between various
    platforms and data centers. Both are well-defined IVOA protocols; especially TAP
    enjoys good software support. Other IVOA recommendations or standards find
    either sporadic use only, or are not considered, for complex reasons. The
    common theme is that robust implementations are missing, and there is insufficient funding to develop them.
    \item Authentication (authN) and authorization (authZ) across centers are
    basically non-existent. Note that providing authN through an established
    protocol (e.g., OAuth) is not the same as mutually federating access
    across science platforms. This is a relatively low-hanging fruit, and
    could be solved if the incentive were to come via strong recommendation
    from the NSF, NASA, etc.
    \item Custom and proprietary data service APIs (if they exist at all)
    make external interaction very challenging. Such APIs are often developed
    at data centers in lieu of more complex standardized interfaces
    (e.g. supporting VO protocols) due to time and funding pressures.
    In some cases VO protocols are initially evaluated. But if they are
    found to have incomplete coverage of all project requirements, they
    are disfavored. The sustainable solution would be instead to engage
    with the IVOA on improving the standards for the data processing needs
    of today. This requires time, and thus stable funding, but the
    benefits would be global.
    \item Combining very large datasets for joint analysis across physical
    locations, temporarily or permanently, is an unsolved problem.
    Proposed solutions range from holding copies of everything at every
    location, over ``data lakes'' that hold copies of everything in just
    one or two locations and are accessible to the entire community, to
    ``streaming'' scenarios where the needed data are federated in
    chunks and temporarily at the place of computation. None of these
    approaches have been successfully implemented yet, and the most
    often practiced solution is to hold local copies of all valuable
    datasets for joint analysis. This is still potentially affordable at
    the level of surveys with a few billion object detections, but not
    at all in the era of Rubin's \gls{LSST} with \textasciitilde 30B
    detected stars and galaxies across up to 1,000 visits.
    \item Shifting (some) data services to commercial cloud providers has a
    two-fold effect. While some aspects of cloud computing have almost
    universally converged to canonical solutions in the industry (e.g.,
    \gls{Docker} for containerization, \gls{Kubernetes} for container
    orchestration, Terraform for infrastructure as code), other components
    of these systems remain hopelessly proprietary (e.g. compute engines,
    file storage mechanisms). Moving between cloud providers is difficult,
    as is sharing components between science platforms (though
    the containerized deployment  makes it much easier than in the past).
    \item The long-term stability of commercially offered cloud services is
    challenging. Services are often replaced with backwards-incompatible
    versions, or dropped entirely. Even ``Enterprise''-level service levels
    often enjoy no more than a year of deprecation warning. For
    relatively small teams at science platforms this can pose
    insurmountable difficulties.
    \item Cost projections for cloud deployments of science platforms
    have almost arcane complexity, making sound and informed choices very
    challenging for the scientific community.
    \item Price pressures may push operations to use proprietary cloud services,
    e.g., BigQuery in favor of an on-premises Postgres database system. Even if
    both support a form of SQL, their syntax may differ, necessitating
    additional development. Even more complex is the case where features
    specific to a single DB engine (e.g., Q3C for Postgres) need
    replacement on another DB; the required research and implementation
    efforts add to the burden on the teams. Similar problems arise when going
    from block-based / POSIX-compatible files storage on-premises, to
    S3 bucket / object store in the cloud (even if commercial translation
    layers do exist).
\end{itemize}

\subsubsection{A model of shared development of science platforms components by design}

We propose that all publicly funded US-based science platforms share
the development of components by design. We do not suggest that a
single science platform is flexible enough to satisfy even a
significant fraction of current science platforms' (sometimes
divergent) requirements. But we argue that science platform developers
and operators should work towards more consolidation of solutions to
common use cases. Consider a model where technical solutions would
exist during planning, development, and beyond, in a ``bazaar of
common components.''  For instance, existing implementations of the
TAP service could find a common foundation through shared development.
One common implementation that is generalized, robust, flexible,
and well tested, could emerge from ``shared development by design.''
Any new operator wishing to put together a science platform-like
system would be empowered to plug \& play interoperable components.

Implementations of new components could be shared very early in
their design and development cycle. If no competing implementations
exist, other data centers and software engineers could lend their
time and expertise to the common cause. All additions and improvements
would be upstreamed by consensus (after any necessary discussions).

A modular framework requires a solid and flexible backplane with clean,
well documented, and stable interfaces. They define how new service
``plugins'' can be added to a system. In the example of a TAP
service, one plugin could handle user queries while individual
``backend'' plugins would handle the underlying database systems
(e.g., Postgres, \gls{Qserv}, etc.). The plugin architecture would
allow additive development.

For every use case it is possible that one technologically ``best''
solution would emerge, but variations of plugins could easily
co-exist in this ecosystem, as long as they all adhere to the ``must be
plugable into the backplane'' paradigm.

Adoption of the backplane \& components model could be gradual.
Different institutions may be already committed to particular technology
stacks, but they can select individual modules when they do not
conflict with the institutional requirements, and they can evolve
to adopt the component model over time. New platforms could
start with this ``standard library'' from the beginning.

We note that commercial solutions exist that are similar in nature
to what we suggest here (e.g. Dataiku), but are likely prohibitively
expensive at scale. Our community should however take such systems
as inspiration lest we continue reinventing the wheel in perpetuity.
Design and implementation of a plugable backplane requires serious
work, but the rewards would be immense.

We finally add that a common-backplane \& plugin components framework
would likely require a canonical deployment mode. Given the continuing
consolidation of version control (\gls{git}), containerization
(\gls{Docker}), orchestration (Kubernetes), continuous integration
and continuous deployment (e.g., ArgoCD), this appears to us to
be the least contentious aspect of the model. A common deployment model
could also easily implement mechanisms to tag pods for deployment on
specific resources, e.g. on GPU clusters for machine learning, or
on high-performance storage farms for massive data processing, etc.

\subsubsection{Where does this fit within the vision of the workshop?}

We stipulate that the proposal would not work on its own without
incentives, without an initial dedicated development effort, and
without staff to maintain the plugin system of code ``exchange.''

Within the structural models explicated in this document, the modular
science platform we describe requires common, collaborative resources to develop the
plugin backplane for a science platform framework. These resources could
be provided either directly as funds to a dedicated project, with PIs
and contributors from all stakeholders in the community. We can also
envision, within the structural model 3 (``The Common Engineering Pool''),
that these engineers could take on the initial task of developing the
backplane framework, in close collaboration with the existing science
platform designers and operators.

Resources would also be needed to continue to maintain and improve the
framework for many years to come, but at a somewhat lower level than the
initial development. Finally, resources for the development or adaptation
of individual plugins should be made available, for instance, on an annual
basis through a competitive grant process. Especially in the Structural
Model 2 (``The Program Office'') a new body would facilitate such funding
rounds, and a review panel  would select the grantees. In the Structure
Model 3 software engineering  resources could be instead requested
(e.g. ``We wish to adapt our existing TAP service to the common backplane'').
This could be either a competitive process or allocated by consensus
(e.g., several proposals for the next plugin development effort to be
funded could be voted on by any and all of the stakeholders).

\pagebreak

%% file: hopes.tex
\section{Hopes and Fears}

The recommendation of this report is to create a new
entity within the U.S. astronomical data system. This recommendation
is potentially transformative, but also carries risks.
When we come together as a community to discuss these issues,
we express hopes for a better future, but we are often held
back by fears that any change will result in a worse
outcome---whether threatening us personally, groups that we may
identify with, or breaking things that we believe are working
well. This is particularly true in an environment where these
discussions are usually held in the context of who gets funding
and who does not.

By explicitly describing the hopes and fears of the participants
of this workshop, we hope that any decision making agencies, the
future body envisaged in this document, and the community at large
can understand the context in which the discussions that resulted
in this article arose, so that it can consider the spirit, as well
as the letter, of the recommendations reached.

\subsubsection{Fears}

\begin{itemize}
    \item We fear that coordination efforts that rely on ``just getting people
    talking'' and hoping that collaboration will naturally arise are
    insufficient, given the existing underlying incentives for
    astronomical \gls{software} development.
    \item We fear that we are not hearing from all the important stakeholders
    and experts.
    \item We fear that differences between the cultures, structures, and
    processes at the three agencies (\gls{DOE}, \gls{NASA} \& \gls{NSF}) will
    hobble a fundamentally cross-agency initiative.
    \item We fear that a body endowed with powers intended to improve
    things would become prescriptive in a way that would squash innovation
    instead of nurturing it.
    \item We fear that funding lines and incentives include gate-keeping that
    is often poorly placed to highly prioritize \gls{software}-centric activities
    such as productization, codebase renewal/refactoring, experimentation and
    generalization to other use cases, since it is often focused on
    specific scientific mission needs that define success.
    \item We fear that entrenched bodies (institutions, researchers,
    collaborations, individuals) may operate exclusively in self-interest, and attempt to
    subvert changes that promote the overall goals but may be disadvantageous
    to their own organization, institution or group.
    \item We fear that organizations and initiatives that are started with
    the best intentions can soon become bureaucratic, self-serving and
    entrenched in the same way as the systems they intended to replace.
    \item We fear that we will once again fail, despite our best efforts,
    to make meaningful progress in this area.
    \item We fear that the balance between perceived innovation and perceived
    stability is difficult to identify in advance even with the best
    intentions.
    \item  We fear that large organized collaborative efforts often result in
    work and resources being allocated inappropriately because of too
    great a focus on political considerations rather than
    technical merit and ability to execute.
    \item We fear that the balance between accountability and overhead is hard
    to strike, and both these extremes are damaging to high velocity but
    responsible progress.
    \item We fear that as a culture we do not have sufficient interest
    and attention available to study well-recognized models for effective
    \gls{software} development for their application in our field.
    \item We fear a Darwinian evolution toward a ``gig economy'' that does
     not value people.
    \item Setting the reasons aside, a substantial number of existing
    interfaces and \gls{software} packages are well behind the state of the art.
    There is the fear that adding free money will be applied to continue the
    same without exploring/encouraging significant improvements/change.
    \item We fear that the astronomical \gls{software} community does not have
    sufficient computer science expertise nor risk tolerance to evaluate and
    test new and emerging technologies on time to implement them for new
    projects.
    \item  We fear that collaborators in the efforts proposed  will not be
    properly incentivized to  promote collective goals.
    \item We fear that the new institute would be “captured” by the existing
    archives, e.g. that instead of addressing the main problems identified in
    this report, the new institute would be a vessel to implement siloed
    priorities for one or more missions or public-facing archives.
    \item We fear that the result of this effort will be a series of unfunded
    requirements that land on individuals and projects and that are not
    achievable.
\end{itemize}

\subsubsection{Hopes}

\begin{itemize}
    \item We hope to encourage international cooperation and that this body
    will effectively and respectfully work across national and cultural
    boundaries.
    \item We hope that our work can support a spirit of joint exploration
    among researchers at all different levels in astronomy.
    \item We hope that we can build a future of vigorous, collaborative
    \gls{software} ecosystems at every level of the end-to-end astronomical \gls{software}
    system.
    \item We hope that we can shift incentives when building the systems
    required for our own mission (be it a single observatory or a
    multi-mission archive) towards the reuse and extension of code written
    by others rather than simply going down our own path.
    \item We hope to increase the number of projects driven by the
    open-source community in astronomy.
    \item We hope to encourage projects inside and outside of astronomy
    to directly self-assess and work toward \gls{software}
    practices that  support better community development.
    \item We hope that a new body can incorporate innovative developments
    that arise in the community, nurture them into services for the
    community, and sustain them.
    \item We hope that we can converge on well-defined and well-conceived
    frameworks and interfaces that allow us both to have multiple options
    for any particular architectural component but also be able to
    evolve and/or replace individual components without being forced out
    of the entire ecosystem.
    \item We hope for a path to a new structure that not only improves the situation for
    specific areas we can currently identify, but creates an overall
    structural environment that incentivizes and renews these goals into the
    unforeseeable future.
    \item We hope that this new ecosystem will enable reproducible and
    replicable science.
    \item We hope that the technical barriers to engagement are outweighed by
    the technical benefits and social support structures we can foster.
    \item We hope that as a field we can remember that people are more
    important than \gls{software}, because it is great people that write great
    \gls{software}.
    \item We hope that a new body can go beyond supporting and sustaining
    \gls{software} for the ecosystem, and can enable and fund new “blue skies” ideas
    to grow the ecosystem.
    \item We hope that we can develop a culture to value documentation,
    maintainability, portability, extensibility, testing and good
    architecture---i.e. best practices---as much as we value function,
    and create opportunities for training in those areas.
    \item We hope that  we will improve the quality and quantity of
    career paths and opportunities within the astronomical data workforce.

\end{itemize}

%% file: costs.tex
\section{Preliminary cost estimates for the structural models} \label{sec:costs}

\subsection{Structural Case Study: Model 1 --- Small Coordinating Committee}

The budget for the coordinating committee would be contingent on the terms of committee appointments (i.e., whether compensated or voluntary). To provide a general estimate, we consider the salaries of administrative and support staff, assuming partial salary support for committee members. The approximate annual budget ranges as follows, in millions of dollars (M\$):

\begin{centering}
\begin{tabular}{l r}\\
\textbf{Administrative Staff:} & \\
3--5  FTE  & \$0.75 -- \$1.5\\
\textbf{Committee:} & \\
5--9 at 25\% FTE  & \$0.25 -- \$1.0\\
\end{tabular}
\end{centering}

\noindent We therefore estimate a total budget between \$1M and \$2.5M per year.
If the committee is based on volunteer effort, the cost of this model is minimal.

\subsection{Structural Case Study: Model 2 --- Inter-agency Program Office} 

The budget for a new program office would hinge on the specific scope and the balance between standard development and funded projects. To provide a rough estimate, we consider factors such as full-time staff and the number of funded projects.

The minimum staff could range from five to ten individuals, with a practical upper limit of twenty to thirty staff, contingent on how much the charter directs the office towards proposing and maintaining standards. Regarding funded projects, we assume an average project duration of three years, targeting a steady-state number of around 10-20 funded programs. However, it's important to note that the budget may scale up or down in accordance with the number of funded programs.

The estimated annual budget ranges as follows, in millions of dollars (M\$):

\begin{centering}
\begin{tabular}{l r}\\
\textbf{Staff:} & \\
5--20   & \$1 -- \$6\\
\textbf{Funded programs:} & \\
8--15 small projects & \$0.5 -- \$5\\
2--5 medium/large projects & \$1 -- \$6\\
overhead / contingency & \$0.5 -- \$1\\
\end{tabular}
\end{centering}

\noindent We therefore estimate a total budget between \$3M and \$18M per year.

\subsection{Structural Case Study: Model 3 --- Common Engineering Pool}

The budget for this model depends on the balance of software engineers, project managers, and administrative staff. 
It also depends on the relative balance of in-house versus ``distributed'' software engineers, who may be partially or fully funded through the institute.
In either case, the budget is entirely staff and institutional operation costs as there is no cash disbursed (only FTE effort).
If we stick with the estimate of 30 in-house software engineers and 15 in distributed positions, this implies 3--5 program managers, and 4--8 admin staff.
The institutional operation costs will depend on whether this is an entirely new entity (requiring a new facility or space) or housed within an existing facility.
For this estimate, we assume the latter.
A rough annual budget ranges as follows, in units of millions of dollars (M\$):

\begin{centering}
\begin{tabular}{l r}\\
\textbf{Staff:}   &  \\
30--40 software engineers and project managers & \$6--\$10 \\
4--8 admin & \$0.5--\$1 \\
Travel and other budget & \$0.5--\$1 \\
Overhead / contingency & \$1--\$2\\
\end{tabular}
\end{centering}

\vspace{0.5em}
\noindent Hence a total budget between \$8M and \$14M per year in steady state.

\subsection{Structural Case Study: Model  4 --- The Centralized Virtual Institute}
The budget for a new federated organization is difficult to estimate, as it builds off of existing projects and institutes.
Federation is inefficient and requires a lot of rules and procedures to achieve its goals.
If this model is considered as, effectively, a community-wide project management office, one might consider:
\begin{itemize}
\item 2--3 project or program managers,
\item 1--2 admin staff,
\item 2--4 system engineers,
\item 1--3 communications officers/staff, and
\item 1--2 business office staff
\end{itemize}
Assuming no actual funds are handled, this virtual institute still employs between eight and fourteen staff members.
A rough annual budget ranges as follows, in units of millions of dollars (M\$):

\begin{centering}
\begin{tabular}{l r}\\
\textbf{Staff:} & \\
8--14   & \$2 -- \$3.5\\
overhead / contingency & \$0.5 -- \$1\\
\end{tabular}
\end{centering}

\noindent We therefore estimate a total budget between \$2.5M and \$4.5M per year.